\documentclass[a4paper,10pt,twocolumn]{article}

\usepackage[utf8]{inputenc}
\usepackage[english]{babel}
\usepackage[normalem]{ulem}
\usepackage{csquotes}
\usepackage{stix}
\usepackage{gensymb}
\usepackage{microtype}
\usepackage{amsmath}
\usepackage{graphicx}
\usepackage{tikz}
\usetikzlibrary{patterns}
\usetikzlibrary{calc,3d}
\usepackage{placeins}
\usepackage[mode=text]{siunitx} 
\usepackage{standalone}
\usepackage{comment}
\usepackage{makecell}
\usepackage{array}
\usepackage{multirow}
\usepackage{rotating}
\usepackage{pdflscape}

\usepackage{adjustbox} 
\usepackage[unicode=true,pdfusetitle,bookmarks=false,bookmarksnumbered=false,bookmarksopen=false,breaklinks=true,pdfborder={0 0 0},pdfborderstyle={},backref=false,colorlinks=false]{hyperref} 

\usepackage[backend=biber,bibencoding=utf8,texencoding=ascii,style=nature,hyperref=true,doi=false,eprint=true,url=false,isbn=false,maxbibnames=30,date=year,abbreviate=true]{biblatex} 

\AtBeginDocument{
  \AtEveryBibitem{\clearfield{issue}\clearfield{eprintclass}}
  \AtEveryCitekey{\clearfield{issue}\clearfield{eprintclass}}
}

\makeatletter 
\AtEveryCitekey{%
  \ifcsundef{blx@entry@refsegment@\the\c@refsection @\thefield{entrykey}}
    {\csnumgdef{blx@entry@refsegment@\the\c@refsection @\thefield{entrykey}}{\the\c@refsegment}}
    {}}
\defbibcheck{onlynew}{%
  \ifnumless{0\csuse{blx@entry@refsegment@\the\c@refsection @\thefield{entrykey}}}{\the\c@refsegment}
    {\skipentry}
    {}}
\makeatother

\usepackage[textwidth=18cm,columnsep=0.4cm,top=18mm,bottom=26mm]{geometry}



\usepackage{caption}
\DeclareCaptionFont{captionsize}{\small} 
\DeclareCaptionLabelSeparator{verticalbar}{~|~}
\DeclareCaptionLabelFormat{abbrev}{Fig.~#2}
\DeclareCaptionLabelFormat{tableabbrev}{Table~#2}
\usepackage[hang,flushmargin]{footmisc} 
\newcommand\blfootnote[1]{%
  \begingroup
  \renewcommand\thefootnote{}\footnote{#1}%
  \addtocounter{footnote}{-1}%
  \endgroup
}

\usepackage[tiny,compact]{titlesec}

\usepackage{lineno}

\newgeometry{textwidth=18cm,columnsep=1cm,top=18mm,bottom=26mm}

\titlespacing*{\section}{0pt}{2.5ex plus 1ex minus .2ex}{0ex plus .2ex}
\titlespacing*{\subsection}{0pt}{2.25ex plus 1ex minus .2ex}{0ex plus .2ex}

\begin{document}

\newrefsegment

\def\dtuelectro{DTU Electro, Department of Electrical and Photonics Engineering, Technical University of Denmark, Ørsteds Plads 343, Kgs.\@ Lyngby, DK-2800, Denmark}
\def\dtuconstruct{DTU Construct, Department of Civil and Mechanical Engineering, Technical University of Denmark, Nils Koppels Allé, Building 404, Kgs.\@ Lyngby, DK-2800, Denmark}
\def\nanophoton{NanoPhoton -- Center for Nanophotonics, Technical University of Denmark, Ørsteds Plads 345A, Kgs.\@ Lyngby, DK-2800, Denmark}

\title{\Large Observation of strong backscattering in valley-Hall photonic topological interface modes}

\def\authormark#1{\textsuperscript{#1}}
\author{\normalsize
  Christian Anker Rosiek\authormark{1,*},
  Guillermo Arregui\authormark{1},
  Anastasiia Vladimirova\authormark{1,2},
  Marcus Albrechtsen\authormark{1},\\
  \normalsize
  Babak Vosoughi Lahijani\authormark{1,2},
  Rasmus Ellebæk Christiansen\authormark{2,3}
  and
  Søren Stobbe\authormark{1,2,$\dagger$}
}
\hypersetup{pdfauthor={Christian Anker Rosiek; Guillermo Arregui; Anastasiia Vladimirova; Marcus Albrechtsen; Babak Vosoughi Lahijani; Rasmus Ellebæk Christiansen; Søren Stobbe}}

\date{\normalsize 16 April 2024}

\twocolumn[
\begin{@twocolumnfalse}
  \maketitle\textbf{%
The unique properties of light underpin the visions of photonic quantum technologies, optical interconnects, and a wide range of novel sensors, but a key limiting factor today is losses due to either absorption or backscattering on defects. Recent developments in topological photonics have fostered the vision of backscattering-protected waveguides made from topological interface modes, but, surprisingly, measurements of their propagation losses were so far missing. Here we report on measurements of losses in the slow-light regime of valley-Hall topological waveguides and find no indications of topological protection against backscattering on ubiquitous structural defects. We image the light scattered out from the topological waveguides and find that the propagation losses are due to Anderson localization. The only photonic topological waveguides proposed for materials without intrinsic absorption in the optical domain are quantum spin-Hall and valley-Hall interface states, but the former exhibits strong out-of-plane losses, and our work therefore raises fundamental questions about the real-world value of topological protection in reciprocal photonics.
\\
  }%
\end{@twocolumnfalse}%
]%
\noindent \blfootnote{
  \authormark{1}\dtuelectro.
  \authormark{2}\nanophoton.
  \authormark{3}\dtuconstruct.
  \authormark{*}\href{mailto:chanro@dtu.dk}{chanro@dtu.dk}
  \authormark{$\dagger$}\href{mailto:ssto@dtu.dk}{ssto@dtu.dk}
}%
Planar nanostructures built with high-index dielectric materials using top-down nanofabrication techniques have enabled precise control of the spatial and spectral properties of electromagnetic fields at optical frequencies, stimulating the development of integrated photonic devices such as quantum light sources~\cite{Lodahl_InterfacingSinglePhotons_2015}, programmable photonics~\cite{Bogaerts_ProgrammablePhotonicCircuits_2020}, nanolasers~\cite{Yu_UltracoherentFanoLaser_2021a}, and optical communication technology~\cite{asghari_energy-efficient_2011}. While highly optimized performance can be achieved with dedicated aperiodic structures~\cite{Albrechtsen_NanometerscalePhotonConfinement_2021a}, periodic structures, i.e., photonic crystals, offer simple building blocks that readily scale to larger architectures and allow tailoring the dispersion relation of light. In addition, deliberately introduced regions that break the periodic symmetry allows building high-$Q$ optical cavities~\cite{Akahane_HighQPhotonicNanocavity_2003} or waveguides that slow light by several orders of magnitude~\cite{Notomi_ExtremelyLargeGroupVelocity_2001}. Introducing such defects can greatly enhance the interaction of light with material degrees of freedom~\cite{Lodahl_InterfacingSinglePhotons_2015,Aspelmeyer_CavityOptomechanics_2014,corcoran_green_2009}.

Long-range translational symmetries can also be effectively destroyed by random structural disorder. This disorder results in extrinsic scattering events, incurring substantial propagation losses~\cite{smith_low-loss_2000} with detrimental consequences for applications. For example, photonic quantum technologies rely on encoding information in fragile quantum states, which are extremely sensitive to losses, and optical interconnects aim to reduce the energy consumption in integrated information technology and here transmission loss translate directly into energy loss~\cite{Hughes_ExtrinsicOpticalScattering_2005a,Lodahl_InterfacingSinglePhotons_2015,Miller_AttojouleOptoelectronicsLowEnergy_2017a}.
Improvements in nanoscale fabrication down to nanometer tolerances can reduce the magnitude of the structural disorder, but stochastic deviations from the designed structures are inherent to any fabrication method and can never be completely eliminated. Since the scattering cross-section is unfortunately often enhanced at the spectral and spatial regions targeted for device operation, such residual disorder is a primary obstacle to the application of photonic crystals~\cite{Hughes_ExtrinsicOpticalScattering_2005a}. A well-known case is that of slow light in photonic-crystal waveguides, where disorder ultimately limits the maximal slowdown by localization of the light field induced by multiple coherent backscattering~\cite{Topolancik_ExperimentalObservationStrong_2007,Sapienza_CavityQuantumElectrodynamics_2010}.

A possible solution to this problem has sought inspiration from solid-state systems, for which the quantum Hall effect offers unidirectional propagation, i.e., completely suppressed backscattering, by breaking time-reversal symmetry. These quantum states are related to the underlying wavevector-space topology of the Bloch eigenstates, which can equally be explored and exploited for photonic-crystal structures, indicating a deep analogy between solid-state quantum states and classical waves~\cite{longhi_quantum-optical_2009,lu_observation_2017}. This naturally led to the development of topological photonics~\cite{Haldane_Analogsofquantum-Hall-effectedgestatesinphotoniccrystals_2008,Ozawa_TopologicalPhotonics_2019} and to the demonstration of one-way robust electromagnetic waveguides~\cite{Wang_ObservationUnidirectionalBackscatteringimmune_2009}, i.e., photonic topological insulator (PTI) waveguides. While early attempts relied on real magnetic fields and non-reciprocal magneto-optical materials to generate non-trivial topologies, further realizations were achieved by effective magnetic fields through time-modulated media~\cite{fang_realizing_2012}. However, such approaches to combat backscattering have seen only microwave-domain implementations that use intrinsically lossy materials~\cite{poo_experimental_2011} or complex active schemes of difficult practical implementation~\cite{minkov_unidirectional_2018}. PTIs that instead rely on breaking spatial symmetries to emulate pesudospins akin to that in quantum spin-Hall (QSH) and quantum valley-Hall (VH) solid-state topological insulators have been predicted~\cite{hafezi_robust_2011,ma_all-si_2016} and demonstrated~\cite{hafezi_imaging_2013,shalaev_robust_2019}. The resulting interface states, albeit not unidirectional owing to reciprocity, can in principle exhibit robustness to a certain class of perturbations~\cite{saba_nature_2020-1}. Both QSH and VH interface states in high-index dielectric photonic-crystal slabs have been observed at telecom wavelengths, but the former support states above the light line~\cite{parappurath_direct_nodate} which are intrisically lossy~\cite{Sauer_TheoryIntrinsicPropagation_2020}, making VH topological interface states particularly attractive to test and exploit topological protection against backscattering.

\begin{figure}[tb]
  \centering
  \includegraphics[width=8.6cm]{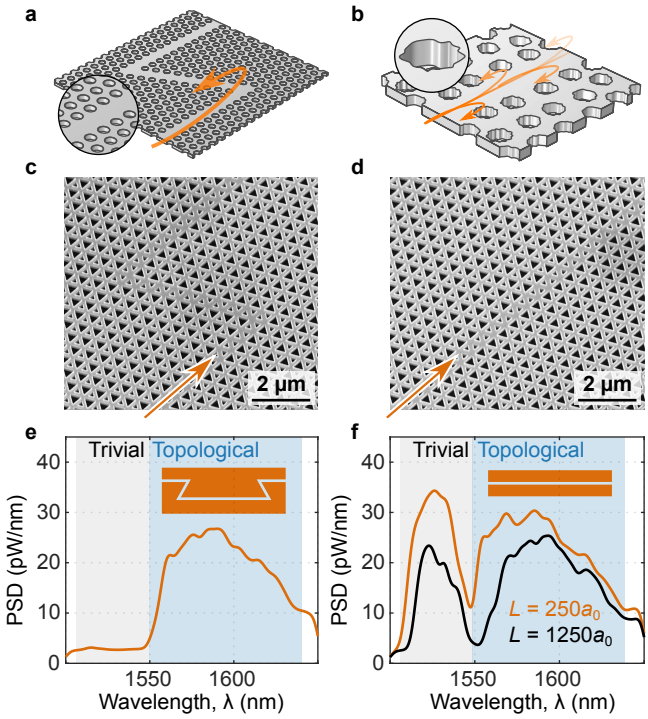}
  \caption{\textbf{Disorder and robustness in photonic-crystal waveguides.}
  \textbf{a}, Illustration of \SI{120}{\degree}-sharp bend, a type of intentionally introduced disorder at the unit-cell scale, in a conventional W1 waveguide.
  \textbf{b}, Illustration of a conventional W1 waveguide with random structural disorder.
  \textbf{c},\textbf{d}, Scanning electron microscopy (SEM) images of a VH PTI waveguide, supporting one topological and one trivial propagating mode, (\textbf{c}) with a sharp bend and (\textbf{d}) without it, i.e., a straight waveguide.
  \textbf{e}, Measured power spectral density (PSD) of light transmitted through the sharp bends, displaying transmission exclusively in the topological band.
  \textbf{f}, Measured PSD of the light transmitted through two PTI waveguides of differing lengths, $L=250a_0$ (orange line) and $L=1250a_0$ (black line), showing the presence of propagation losses. Comparing \textbf{e} and \textbf{f} indicates that while topological modes can propagate around sharp bends, they do not offer significant protection against backscattering from real-world disorder.}
  \label{fig:1}
\end{figure}

\section*{Disorder length-scales in photonic-crystal structures}
\noindent Much of the existing work on disorder in topological photonics has explored topological protection against defects on scales relevant to their electronic counterpart. Topological quantum states of electrons can travel unhindered along paths prone to crystallographic defects such as vacancies, interstitials, or dislocations~\cite{ni_robustness_2020}, which are all on the scale of one to a few crystal unit cells. Such lattice-scale disorder has been mimicked in PTIs, with the most paradigmatic case being that of sharp Z- or $\Omega$-shaped bends. Unlike in conventional line-defect photonic-crystal waveguides as that shown in Fig.~\ref{fig:1}a~\cite{Arora_DirectQuantificationTopological_2021}, suppressed back-reflection through sharp bends over a large bandwidth has been demonstrated for topological interface states ~\cite{shalaev_robust_2019,Yoshimi_ExperimentalDemonstrationTopological_2021b}, enabling flexibly-shaped photonic circuits like ring cavities~\cite{Xie_TopologicalCavityBased_2021}. Nanophotonic waveguides are, however, prone to nanometer-scale roughness in the etched sidewalls as exemplified in Fig.~\ref{fig:1}b. Such structural disorder occurs at a scale significantly smaller than the unit cell and is consistenly present across the entire crystal, thus questioning how the notions of topological protection can be directly transferred to the interface states in PTIs. Recent numerical works have addressed this question, but the studies are limited to effective disorder models in two-dimensional crystals~\cite{Arregui_QuantifyingRobustnessTopological_2021}, semi-analytical models~\cite{Orazbayev_QuantitativeRobustnessAnalysis_2019}, or single-event incoherent scattering theory~\cite{hauff_chiral_2022}. More accurate modelling of the propagation losses including coherent multiple scattering has been applied to conventional monomode slow-light waveguides~\cite{Patterson_DisorderInducedCoherentScattering_2009,Mazoyer_DisorderInducedMultipleScattering_2009}, but such studies are still lacking for topological interface states. Ultimately, the subtle interplay between out-of-plane losses and backscattering in photonic-crystal slab waveguides calls for experiments that directly compare the propagation losses of topological and conventional slow-light waveguides subject to equivalent disorder while taking into account the group index. Here, we address this by fabricating and characterizing a set of suspended silicon VH PTI waveguides~\cite{Yoshimi_SlowLightWaveguides_2020} that support both a topologically protected and a topologically trivial guided mode with nearly identical group indices. We characterize waveguides with and without sharp bends, cf.\ Figs.~\ref{fig:1}c and \ref{fig:1}d. As shown in previous work~\cite{Yoshimi_ExperimentalDemonstrationTopological_2021b}, the difference between the two guided modes is profoundly manifest when introducing four sharp turns in the waveguide path. They effectively suppress the transmittance in the trivial mode, while leaving transmission through the topological mode essentially unaffected as shown in Fig.~\ref{fig:1}e. This is a striking demonstration of topological protection, but it offers no evidence of the applicability of PTIs for protecting against backscattering from fabrication imperfections. It is clear from simple experiments, e.g., comparing the transmittance of short and long waveguides as shown in Fig.~\ref{fig:1}f, that the propagation loss of the topological mode studied here is nonzero. However, assessing topological protection against backscattering requires the precise extraction and modeling of the propagation losses to disentangle out-of-plane radiation losses from backscattering, the careful study of which has so far been absent from literature.

\begin{figure}[tb]
  \centering
  \includegraphics[width=8.6cm]{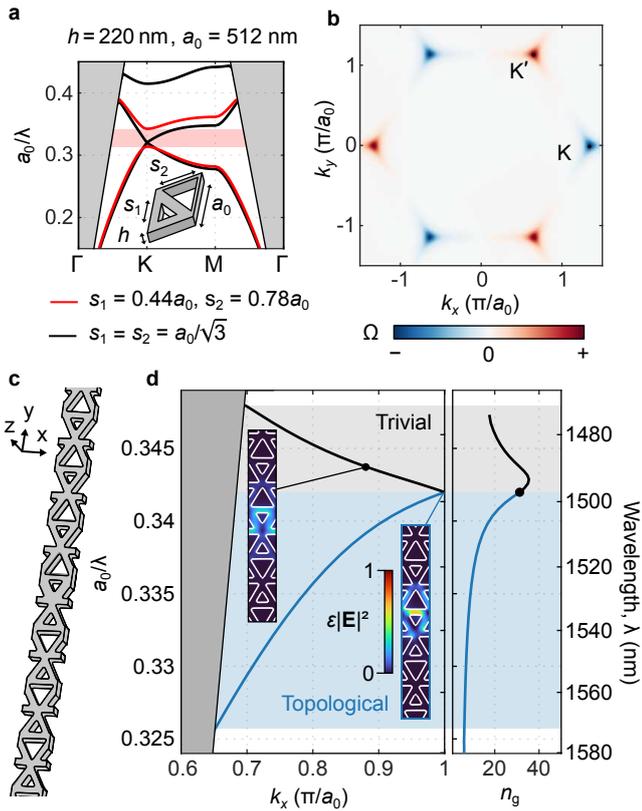}
  \caption{\textbf{Valley-Hall photonic crystals and interface states.}
  \textbf{a}, Band diagram for valley-Hall (VH) photonic crystals (PhCs) with (black) and without (red) inversion symmetry. The inset shows the two-triangle unit cell. The band gap generated at the K-point by breaking the inversion symmetry ($s_1<s_2$) is indicated in shaded red.
  \textbf{b}, Berry curvature $\Omega$ for the dielectric band in {\bfseries a}.
  \textbf{c}, Schematic of the supercell generated by interfacing two VH PhCs using a glide-symmetry operation. The silicon geometry (displayed in gray) is surrounded by air.
  \textbf{d}, Dispersion diagram and group index for the structure shown in {\bfseries c}. The insets show the electrical energy density, $\varepsilon|\mathbf{E}|^2$, for the trivial (black solid line and gray shading) and topological (blue solid line and blue shading) Bloch modes at a similar group index of $n_\text{g}=31$ with the outlines of the geometry overlayed. For the topological mode, this group index occurs at the symmetry-protected degeneracy point, $k_x=\pi/a_0$. \label{fig:2}
  }
\end{figure}

\begin{figure*}[t]
  \centering
  \includegraphics[width=17.8cm]{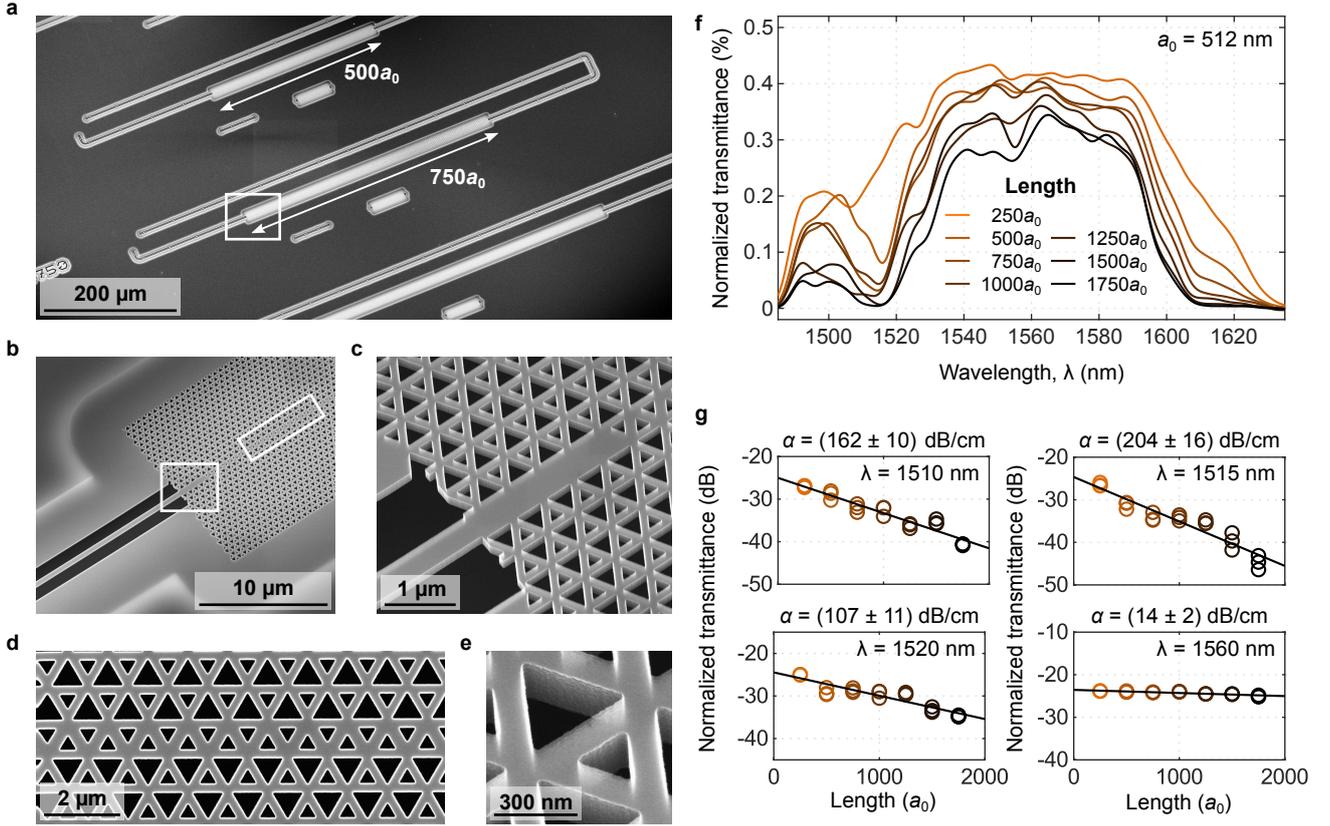}
  \caption{\textbf{Experimental characterization of propagation losses.}
  \textbf{a}, Scanning electron microscopy (SEM) image of part of an array of photonic topological insulator (PTI) waveguides of lengths 250, 500, 750, 1000, 1250, 1500 and 1750 unit cells embedded in otherwise equivalent photonic circuits.
  \textbf{b},\textbf{c}, Close-up views of coupling between strip and PTI waveguides.
  \textbf{d}, Top-down view of a section of PTI waveguide.
  \textbf{e}, High-magnification SEM image showing a single unit cell of the PTI.
  \textbf{f}, Transmittance data for one full array of test circuits of lengths from $250a_0$ to $1750a_0$ (see legend).
  \textbf{g}, Length-dependent logarithmic transmittance at four representative wavelengths. All three nominally identical devices for each waveguide length are included and the solid black lines show the least-square fit from which the loss length is obtained. Color indicates length as in \textbf{f}. \label{fig:3}}
\end{figure*}

\section*{Design of a valley-Hall slow-light waveguide}
\noindent The VH PTI waveguides explored here rely on a photonic crystal that emulates graphene with two equilateral triangular holes arranged in a honeycomb structure, the unit cell of which is shown in the inset of Fig.~\ref{fig:2}a. 
For identical triangles, $s_1=s_2$, the crystal dispersion exhibits a Dirac cone for transverse-electric-like (TE-like) modes at the $\mathrm{K}$-point in the Brillouin zone. Breaking inversion symmetry such that $s_1\neq s_2$ lifts the degeneracy and opens a band gap, as shown in Fig.~\ref{fig:2}a. The non-trivial geometrical structure in momentum space of the wave functions of the air and dielectric bands of the crystal results in non-vanishing Berry curvatures~\cite{Wang_UniversalNumericalCalculation_2020} around the $\mathrm{K}$ and $\mathrm{K}'$ points, as shown in Fig.~\ref{fig:2}b. When two such crystals with inverted symmetry are interfaced, the bulk-edge correspondence theorem~\cite{ma_all-si_2016} ensures the existence of two degenerate counterpropagating states localized at the domain wall and exhibiting a linear dispersion around the projection of K in the waveguide direction. The particular geometry explored here (Fig.~\ref{fig:2}c) uses a bearded interface~\cite{Ozawa_TopologicalPhotonics_2019} between two mutually inverted VH crystals. In comparsion to the commonly investigated zigzag interface~\cite{shalaev_robust_2019}, the bearded interface obeys a composition of mirror and translation symmetry, i.e., a glide symmetry. This enforces a degeneracy at the edge of the Brillouin zone~\cite{Mock_SpaceGroupTheory_2010}, leading to the existence of two guided modes (Fig.~\ref{fig:2}d). The superior transmission through sharp bends at wavelengths in the low-energy band relative to those in the high-energy band, shown in Fig.~\ref{fig:1}e, indicates that the former is topological (see Supplementary Section 6), and shows that protection to back-reflection at sharp 120\degree{} bends extends far from the K-valley. The dispersion diagram is computed from geometric contours extracted from scanning electron microscopy (SEM) images (see Supplementary Section 1.2) and shows that the fabricated structure is single-moded (cf. Supplementary Section 7). Figure~\ref{fig:2}d also shows the group index as a function of wavelength. A group index of $n_{\text{g}}\sim 30$ is achieved in the topological band, a value for which backscattering typically dominates out-of-plane radiation losses~\cite{Mazoyer_DisorderInducedMultipleScattering_2009}, making the waveguide an ideal testbed to study the aforementioned topological protection to backscattering.

\section*{Characterization of optical propagation losses}

\noindent We characterize the dispersive propagation losses of suspended silicon photonic-crystal waveguides fabricated with a slab-thickness of \SI{220}{nm}. This is achieved by measuring the optical transmission of suspended photonic circuits where waveguides of varying length, $L$, from $250a_{\text{0}}$ to $1750a_{\text{0}}$, with $a_{\text{0}} = \SI{512}{nm}$ denoting the lattice constant, are embedded. Scanning electron microscopy images of characteristic devices are shown in Fig.~\ref{fig:3}a. The circuits are comprised of input and output free-space broadband grating couplers, strip silicon waveguides to direct light into the region of interest (Fig.~\ref{fig:3}b), and intermediate waveguides~\cite{Chen_Efficient_2022} (Fig.~\ref{fig:3}c) to couple to the VH interface modes (Fig.~\ref{fig:3}d) with high efficiency (\SI{87}{\percent}, see Supplementary Section 2.3). Figure~\ref{fig:3}e shows a high-magnification SEM image, which reveals the presence of roughness along the sidewalls. In principle, the roughness could be measured and used to calculate the scattering but, in practice, such a procedure is experimentally unfeasible and numerically intractable 
\cite{le_thomas_statistical_2011,Arregui_QuantifyingRobustnessTopological_2021}. Therefore, we instead benchmark our VH waveguides against conventional line-defect W1 waveguides fabricated on the same chip such that the structural disorder has practically equivalent statistics. We extract an average propagation loss as low as \SI{0.47+-0.04}{dB/cm} over a \SI{40}{nm} bandwidth in the non-dispersive region of the W1-waveguide (see Supplementary Section 3.2). This constitutes a record-low value for suspended silicon photonics and shows that our nanofabrication~\cite{Albrechtsen_NanometerscalePhotonConfinement_2021a} provides an ideal testing ground for measuring VH-waveguides with the lowest level of roughness realized to date.

The circuit transmittance for a single VH device for each waveguide length is shown in Fig.~\ref{fig:3}f. We convolute the raw spectra with a Gaussian kernel (standard deviation $\sigma = \SI{2.5}{nm}$) to simultaneously remove Fabry-Pérot fringes resulting from reflections at the grating couplers and account for possible systematic structure-to-structure variations (see Supplementary Section 2.1). Inside the transmission band, we observe that the loss is largest around $\lambda=\SI{1515}{nm}$, which is a clear spectral indication of the $n_{\text{g}}$-peak shown in Fig.~\ref{fig:2}d. We account for the stochastic nature of the sidewall roughness by studying averaged quantities obtained from measurements done over 3 nominally identical circuits for each waveguide length. We find that the ensemble-averaged transmission intensity can be described by an exponential spatial decay, with an attenuation coefficient $\alpha(\lambda)\equiv\ell_L^{-1}(\lambda)$. This leads to the following damping law
\begin{equation}
 \label{eq:logT-eq}
 \langle \ln T(\lambda) \rangle = -{L\over \ell_L(\lambda)} + \ln T_0(\lambda)
\end{equation}
where additional losses in the circuit are cast into $T_{\text{0}}$ (see Supplementary Section 2.1). Characteristic fits to the ensemble-averaged data are shown in Fig.~\ref{fig:3}g.
While such Beer-Lambert-like attenuation has been theoretically shown to fail for particular periodic monomode waveguides~\cite{baron_attenuation_2011}, the moderate values of $n_{\text{g}}$ explored here and the state-of-the-art nanofabrication process justify the model. The same arguments also support the use of 3 devices per length since the variance of the stochastically distributed transmission, which depends on the loss pathway, group index, and waveguide length~\cite{Mazoyer_DisorderInducedMultipleScattering_2009}, is low as confirmed by data subsets with increasing number of nominally identical devices (see Supplementary Section 2.2).

\begin{figure}
  \centering
  \includegraphics[width=8.6cm]{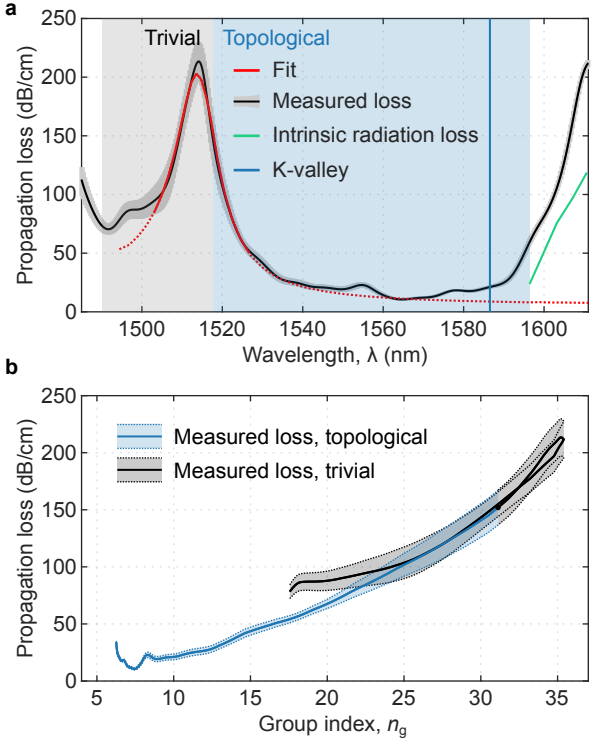}
  \caption{\textbf{Dispersive propagation losses in a slow-light glide-symmetric valley-Hall waveguide.}
  \textbf{a}, Wavelength dependence of the measured propagation loss (solid black line) obtained from the fit coefficients shown Fig.~\ref{fig:3}g, with the shaded area indicating the standard error. The propagation loss is fitted with the model in Eq.~\eqref{eq:loss-model} (red line) and the trivial-topological transition shaded according to the found wavelength offset. The solid segments of the fit line indicate the intervals that were fitted. The calculated intrinsic loss above the light line is included for reference (solid green line). The location of the K-valley in the topological mode is indicated by a vertical blue line.
  \textbf{b}, Dependence of the loss on the group index, $n_{\text{g}}$, where the wavelength to group-index conversion is derived from the fit in {\bfseries a}.\label{fig:4}
  }
\end{figure}

\begin{figure*}
  \centering
  \includegraphics[width=17.8cm]{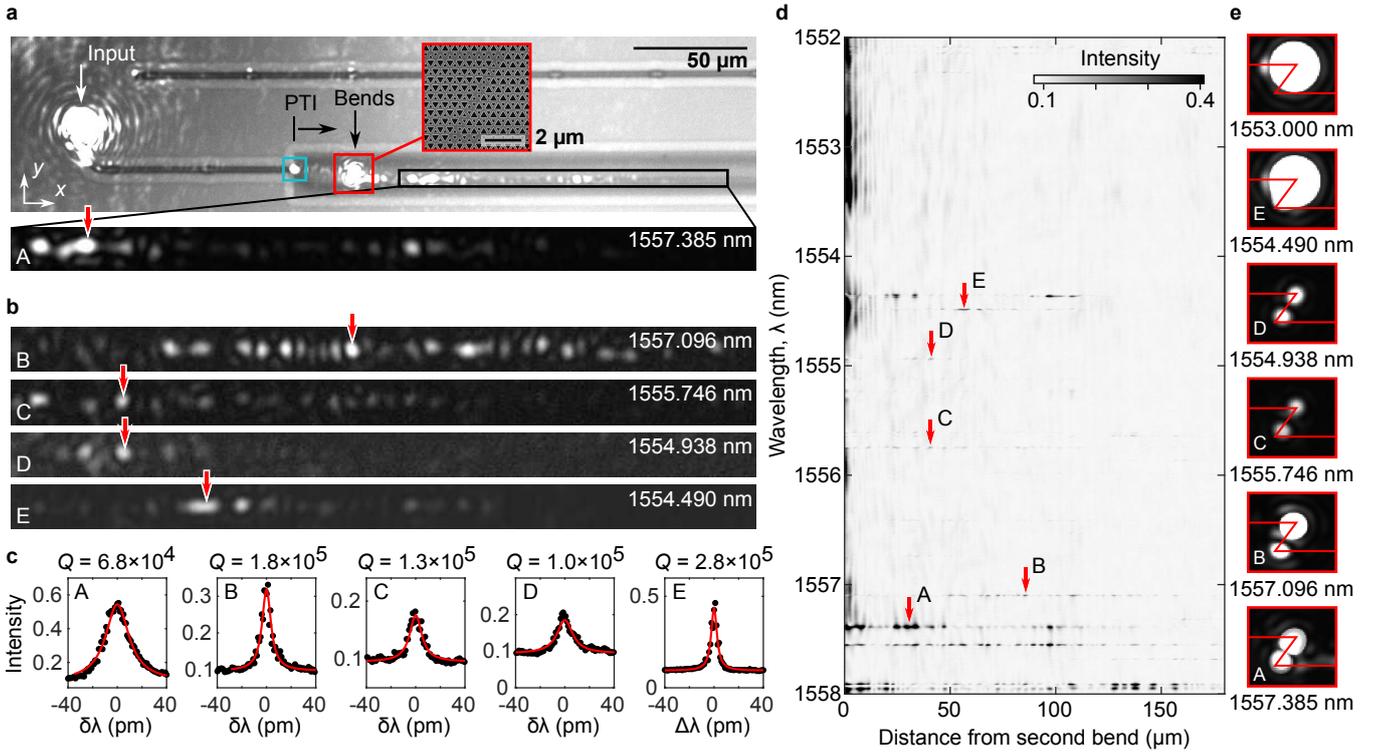}
  \caption{\textbf{Observation of coherent backscattering in a valley-Hall topological interface state.} \textbf{a}, Microscope image of a photonic circuit and overlayed vertically scattered far-field at a wavelength $\lambda = \SI{1557}{nm}$. Light polarized in the $x$-direction is excited at the input grating coupler (indicated) and guided to a PTI waveguide. The cyan box highlights the scattering site at the strip-to-intermediate waveguide interface and the red box the scattering sites in the sharp bend. The black box highlights the scattered fields along the waveguide axis after the second corner, a detail of which (A) is given below.
  \textbf{b}, A representative set of far-fields (labeled B--E) at wavelengths corresponding to high-$n_{\text{g}}$ values of the underlying topological interface state.
  \textbf{c}, Spectral resonances associated with modes A--E obtained by averaging over a small set of pixels near the corresponding arrows in {\bfseries b}. The horizontal axis shows detuning from the resonance wavelengths, $\delta\lambda$.
  \textbf{d}, Spatial and spectral mapping of the scattered far-fields after the second corner and along the waveguide axis. The wavelength range corresponds to the region around the degeneracy point. The modes shown in {\bfseries b} are highlighted.
  \textbf{e}, Detailed view of the scattered fields at the vertices of the bend corresponding to modes A--E and to a wavelength well within the topologically trivial band. Light propagating in the trivial mode is lost from the waveguide due to strong scattering at the first bend while the topological mode offers protection from scattering at the bends such that scattering from both bends can be observed, which allows a clear distinction between trivial and topological modes. \label{fig:5}}
\end{figure*} 

\section*{Group-index dependence}
The propagation loss over a wide wavelength range is shown in Fig.~\ref{fig:4}a. It exhibits a prominent dispersion across the maximal group index around $\lambda = \SI{1515}{nm}$. As a consequence of the strong dispersion, the extracted propagation loss depends on the width of the filtering kernel so the values shown in Fig.~\ref{fig:4}a constitute a lower bound to the propagation losses. Broadly speaking, the dependence of the propagation loss on wavelength reflects the physics underlying several distinct scattering mechanisms~\cite{Kuramochi_DisorderinducedScatteringLoss_2005}. The losses of a propagating mode in a photonic-crystal waveguide can be classified into intrinsic losses, $\ell_i^{-1}(\lambda)$, scattering into radiation modes in the cladding, $\ell_\text{out}^{-1}(\lambda)$, and inter- or intramodal scattering into other slab or waveguide modes. Intrinsic losses include absorption material losses, which can be neglected in crystalline silicon at telecom wavelengths except for two-photon absorption in ultra-high $Q$ cavities~\cite{barclay_nonlinear_2005}, and intrinsic radiation losses when operating above the light line~\cite{Sauer_TheoryIntrinsicPropagation_2020}. In the case of a monomode photonic-crystal waveguide with vertical sidewalls operated at wavelengths within the bulk band gap, all sources of inter- and intramodal scattering except for backscattering, $\ell^{-1}_s(\lambda)$, are strongly suppressed. This holds provided that the disorder levels are perturbative and that the number of unit cells in the direction perpendicular to the waveguide axis, here 16, significantly exceeds the Bragg length. All such conditions are satisfied for the VH waveguides explored here, and the remaining loss lengths add up reciprocally to the propagation length,
\begin{equation}
\label{eq:loss-channels}
\ell_L^{-1}(\lambda) = \ell_s^{-1}(\lambda) + \ell_{\text{out}}^{-1}(\lambda)\,.
\end{equation}
Both contributions are generally dispersive and may vary independently based on geometry, material properties, disorder, and wavelength. As a consequence, the precise scaling with $n_{\text{g}}$ is far from trivial~\cite{Mazoyer_DisorderInducedMultipleScattering_2009}. Nevertheless, based on the experimental observations and perturbation theory~\cite{Kuramochi_DisorderinducedScatteringLoss_2005}, we model $\ell_L^{-1}(\lambda)$ by
\begin{equation}
\label{eq:loss-model}
\ell_L^{-1}(\lambda)= \beta n_{\text{g}}^2(\lambda-\Delta\lambda) + \gamma n_{\text{g}}(\lambda-\Delta\lambda)
\end{equation}
where $n_{\text{g}}(\lambda)$ is the theoretical group index shown in Fig.~\ref{fig:2}d. The coefficients $\beta$ and $\gamma$ describe the loss due to backscattering and out-of-plane radiation. Furthermore, to account for the observed spectral shifts of about \SI{20}{nm} between the calculated $n_{\text{g}}$ and the observed loss peak, we introduce an additional model parameter, $\Delta\lambda$, describing a linear spectral shift between theory and experiment. The fit is shown in Fig.~\ref{fig:4}a and agrees well with the experiment, which shows that Eq.~\eqref{eq:loss-model} describes the measured losses well, irrespective of the transition between topological and trivial modes. The fitted coefficients are $\beta=\SI{0.151 +- 0.003}{dB/cm}$ and $\gamma=\SI{0.39 +- 0.07}{dB/cm}$. The band structure is shifted by $\Delta\lambda=\SI{20.22 +- 0.05}{nm}$, which accounts for minor deviations in average dimensions between model and samples, and this shift is henceforth applied to all theoretical quantities, placing the degeneracy point at $\SI{1517}{nm}$ and the K-valley of the topological mode at $\SI{1587}{nm}$
We note that the measured propagation losses are not minimal at the K-valley. For reference, the propagation loss was also measured in both zigzag and bearded interface waveguides using a different unit cell~\cite{shalaev_robust_2019} (see Supplementary Section 3.3). The minimal propagation loss does not coincide with the location of the K-valley for any of the measured waveguides.
The calculated intrinsic radiation losses (see Methods) of the topological mode above the light line is included for reference in Fig.~\ref{fig:4}a and shows good agreement with the measured propagation loss. We obtain the propagation loss as a function of group index shown in Fig.~\ref{fig:4}b and observe that the losses of the two modes coincide within the statistical uncertainty. In addition, we use the fitted $\gamma$ and $\beta$ to infer that, similar to conventional W1 photonic-crystal waveguides~\cite{Mazoyer_DisorderInducedMultipleScattering_2009}, backscattering losses exceed out-of-plane radiation losses even at very low group indices which include the vicinity of the K-valley. 
We conclude that the topological interface mode incurs equivalent propagation losses as the trivial mode in the slow-light regime and thus we observe no topological protection against fabrication disorder.

\section*{Observation of coherent backscattering}
We now turn to exploring the physics of backscattering in the topological waveguide beyond the ensemble behaviour. We image the vertically scattered far-fields from single waveguide realizations ($L=1750a_{\text{0}}$) using a tunable laser and a near-infrared camera. These measurements use a different sample with propagation losses comparable to those reported in Fig.~\ref{fig:3} (see Supplementary Section 3.2). Since it is challenging to distinguish between the topological and the trivial mode from far-field measurements on straight waveguides (see Supplementary Section 4), we employ sharply-bent waveguides, which act as highly efficient (see Supplementary Section 5) filters that only allow the topological mode to pass~\cite{Yoshimi_SlowLightWaveguides_2020}. Figure~\ref{fig:5}a shows a microscope image of a device, overlayed with the scattered far-field at a wavelength well within the topological band. In addition to scattering at the interface between the strip waveguide and the intermediate waveguide and at the two vertices of the bend, we observe spatially varying scattering in a finite region of the waveguide close to the second corner. This is a clear fingerprint of strong coherent backscattering leading to complex interference patterns in the near-field of the propagating mode and projected into the far-field by out-of-plane radiation losses~\cite{Topolancik_ExperimentalObservationStrong_2007}. Coherent backscattering ultimately leads to spectral and spatial localization of the light field and one-dimensional Anderson localization~\cite{Sapienza_CavityQuantumElectrodynamics_2010}. In a single realization, this corresponds to the formation of random optical cavities with distinct spatial patterns (Fig.~\ref{fig:5}b) and quality factors reaching $Q \simeq 2\cdot10^5$ (Fig.~\ref{fig:5}c). The observation of random Anderson-localized cavities with high quality factors corroborates that coherent backscattering is the dominant source of loss at the imaged wavelengths, i.e., that the inverse loss length reported in Fig.~\ref{fig:4} may be interpreted as the inverse localization length~\cite{savona_electromagnetic_2011}.
Finally, we step the laser wavelength across a wide wavelength range around the degeneracy point and study the transition from the topological to the trivial mode. The spectro-spatial map in Fig.~\ref{fig:5}d depicts the acquired intensities after the second corner and along the axis of the waveguide. It reveals the presence of multiple spectral resonances associated with spatially localized far-field patterns, whose spatial extent generally increases with wavelength, as expected from the behaviour of the ensemble-averaged loss length shown in Fig.~\ref{fig:4}. The modes labelled A--E correspond to the images in Figs.~\ref{fig:5}a and b and are not visible in the transmission spectrum of the circuit, which further evidences their localized nature. In addition, a close-up of the bend (Fig. ~\ref{fig:5}e) unveils the topological or trivial nature of the propagating modes. For all wavelengths below $\lambda = \SI{1554}{nm}$, we observe a single scattering spot at the first corner (Fig.~\ref{fig:5}e) and no emission from the waveguide (Fig.~\ref{fig:5}d), indicating wavelengths within the trivial band, consistent with previous observations for this waveguide geometry~\cite{Yoshimi_ExperimentalDemonstrationTopological_2021b}.
At the resonant wavelength of the mode E, light is also scattered strongly at the first corner.
This is consistent with our numerical simulations of high-group-index wavelengths near the degeneracy, even in the topological band (see Supplementary Section 5).
At the wavelengths of the modes labelled A--D, the radiation losses in both corners are not only suppressed but also evenly distributed, confirming the existence of localized modes within the topological band.

\section*{Conclusion and outlook}
The experiments presented here, both the dependence of the measured loss length on the group index as well as the scattered light observed via far-field imaging, establishes a consistent picture of the transmission and scattering of slow light in VH PTI interface modes: Backscattering dominates over out-of-plane losses and is sufficiently strong to induce random cavities with high quality factors. Additionally, we observe no difference between the dependence of the loss length on group index for topological and trivial modes.

Obviously, our experiments do not rule out the existence of backscattering resilience in other time-reversal-invariant PTIs with different symmetries, unit-cell geometries, interfaces, or disorder levels. Even if structural disorder eventually destroys the crystal symmetry behind non-trivial topologies, backscattering might still be suppressed for limited disorder~\cite{Arregui_QuantifyingRobustnessTopological_2021}. To approach that regime, we have employed a bearded VH interface, which has been theoretically shown to be more robust than other types of interfaces~\cite{Orazbayev_QuantitativeRobustnessAnalysis_2019}. In addition, our record-low-loss W1 waveguides show that we are probing the lowest levels of disorder realized in silicon photonics so far. Even so, we do not observe any signature of reduced backscattering and our results therefore cast doubts on whether any topological protection against backscattering from nanoscale roughness is possible in an all-dielectric platform, i.e., without breaking time-reversal symmetry or dynamic modulation techniques.

We hope that our work will motivate further research to consider robustness against real-world disorder, e.g., in developing new magneto-optic materials to break time-reversal symmetry at optical frequencies~\cite{guglielmon_broadband_2019} or studying the mechanisms behind Anderson localization~\cite{Garcia_TwoMechanismsDisorderinduced_2017} in systems with valley-momentum locking. The interplay between disorder and topology has surprising consequences, such as topological Anderson insulators~\cite{Stutzer_PhotonicTopologicalAnderson_2018}, and our work takes the first steps into research at the nexus between Anderson localization, topology, and silicon photonics.

\section*{Acknowledgments}
We gratefully acknowledge Ali Nawaz Babar and Thor August Schimmell Weis for assistance with the nanofabrication and valuable discussions and Mathias Linde Korsgaard for assistance with device design.
The authors gratefully acknowledge financial support from the
Villum Foundation Young Investigator Programme (Grant No. 13170),
Innovation Fund Denmark (Grant No. 0175-00022 -- NEXUS and Grant No. 2054-00008 -- SCALE),
the Danish National Research Foundation (Grant No. DNRF147 -- NanoPhoton),
Independent Research Fund Denmark (Grant No. 0135-00315 -- VAFL), the European Research Council (Grant. No. 101045396 -- SPOTLIGHT), and the European Union's Horizon 2021 research and innovation programme under a Marie Sklodowska-Curie Action (Grant No. 101067606 -- TOPEX).

\section*{Author contributions}
C.A.R., G.A., S.S. designed and developed the experiment.
C.A.R., G.A., A.V., M.A., B.V.L. performed the numerical design and analysis of the structures and device components.
M.A. developed the nanofabrication process.
C.A.R. fabricated the samples.
C.A.R. and G.A. carried out the measurements and data analysis.
C.A.R., G.A., A.V., and S.S. prepared the manuscript with input from all authors.
S.S. conceived, initiated, and supervised the project with co-supervision by G.A. and R.E.C.

\section*{Competing interests}
The authors declare no competing interests.

\printbibliography[segment=\therefsegment,check=onlynew] 

\section*{Methods}

\small

\newrefsegment

\paragraph{Sample fabrication.}
The measurements are performed on two samples denoted Sample 1 and Sample 2. The data in Figs.~\ref{fig:1}~and~\ref{fig:5} are taken from Sample 1 and the data shown in Figs.~\ref{fig:2}--\ref{fig:4} are from Sample 2. Both samples are fabricated from the same silicon-on-insulator substrate with a nominally \SI{220}{nm}-thick silicon device layer. The fabrication process is detailed in Ref.~\cite{Albrechtsen_NanometerscalePhotonConfinement_2021a} with some minor modifications. Sample 1 is fabricated using a high-resolution electron beam lithography process, the details of which may be found in Ref.~\cite{Florez_EngineeringNanoscaleHypersonic_2022}. Sample 2 is fabricated using a modified process, which introduces a silicon-chromium hardmask, following Ref.~\cite{Arregui_CavityOptomechanicsAndersonlocalized_2022}.

\paragraph{Optical spectral measurements.}
The transmission of each device is measured using a confocal free-space optical setup with cross-polarized and spatially offset excitation and collection achieved via orthogonal free-space grating couplers. The broadband optical characterization is performed using a fiber-coupled supercontinuum coherent white-light source (NKT Photonics SuperK EXTREME EXR-15) focused onto the input grating coupler using a long-working-distance apochromatic microscope objective (Mitutoyo Plan Apo NIR 20X, NA = 0.4, \SI{10}{mm} effective focal length). The input power (typically \SI{120}{\micro W} at the sample surface) is controlled using a half-wave plate and a polarizing beamsplitter and the excitation polarization selected with a half-wave plate. Light coupled out from the chip is collected using the same microscope objective, split via a 50:50 beamsplitter and filtered in polarization and space, respectively, using a linear polarizer and a single-mode fiber aligned to the output grating coupler. The light is then sent to an optical spectrum analyzer (Yokogawa AQ6370D, \SI{2}{nm} resolution bandwidth) for retrieving the spectrum. When shown dimensionless, transmittance spectra are normalized to reference measurements on a silver mirror (Thorlabs PF10-03-P01) substituted in place of the fabricated sample (see Supplementary Section 2.1). For imaging the vertically scattered fields, we employ the same free-space optical setup but use instead a fiber-coupled tunable diode laser (Santec TSL-710) for excitation and focus the scattered light into a near-infrared/visible camera (Aval Global ABA-013VIR) using a long-focal-length ($f = \SI{200}{mm}$) plano-convex lens. The intense direct reflection from the input grating coupler is filtered out using a linear polarizer perpendicular to the waveguide axis in the collection path. The camera also serves the purpose of imaging the sample surface using a near-infrared light-emitting diode ($\lambda$ = $\SI{1.2}{\micro m}$).

\paragraph{Numerical modeling.}
We employ a finite-element method using commercially available software (COMSOL Multiphysics) for the numerical calculations. Silicon is modeled as a lossless dielectric with refractive index $3.48$ and air with a refractive index of $n=1$. We simulate the optical eigenmodes of the perfect photonic-crystal waveguides by terminating them with perfectly matched layers in the $y$ and $z$ directions and imposing Floquet boundary conditions (BCs) on both facets of the supercell along the waveguide axis, $x$. The real part of the eigenfrequencies $\omega_{k}$ are used for the band structures in Fig.~\ref{fig:1} and the imaginary part to extract the intrinsic losses above the light cone shown in green in Fig.~\ref{fig:3}a. The latter are obtained as $\ell_{\text{i,dB}}^{-1} = 4.34\times\big(2\operatorname{Im}(\omega_{\textbf{k}})/|v_{\text{g}}|\big)$, with $v_{\text{g}}$ the group velocity.
The transmission simulations shown in Supplementary Sections 2.3 and 5 are solved as frequency-domain problems using the fundamental strip waveguide mode for both input and output ports. All simulations use the symmetry relative to the mid-plane of the slab and solve for transverse-electric-like electromagnetic fields. This relies on the assumption of vertical sidewalls, which we observe, as well as the absence of scattering into transverse-magnetic-like modes.

\section*{Data availability}
The data that support the figures in this manuscript are available from the corresponding author upon request.

\section*{Code availability}
The code that support the figures in this manuscript are available from the corresponding author upon request.

\printbibliography[segment=\therefsegment,check=onlynew]
\endrefsegment

\clearpage
\normalsize

\begin{refsection}

\newrefcontext[labelprefix=S]
\setcounter{equation}{0}
\setcounter{figure}{0}
\setcounter{table}{0}
\setcounter{page}{1}
\renewcommand{\thepage}{S\arabic{page}}
\pagestyle{plain}
\renewcommand{\theequation}{S\arabic{equation}}
\renewcommand{\thefigure}{S\arabic{figure}}
\renewcommand{\thetable}{S\arabic{table}}
\captionsetup[figure]{font=captionsize,labelformat=abbrev,labelfont=bf,labelsep=colon}
\captionsetup[table]{font=captionsize,labelformat=tableabbrev,labelfont=bf,labelsep=colon}

\titleformat{\section}{\normalfont\large\bfseries\filcenter}{\thesection}{1em}{}
\titleformat{\subsection}{\normalfont\normalsize\bfseries\filcenter}{\thesubsection}{1em}{}
\titlespacing*{\section}{0pt}{3.5ex plus 1ex minus .2ex}{2.3ex plus .2ex}
\titlespacing*{\subsection}{0pt}{3.25ex plus 1ex minus .2ex}{1.5ex plus .2ex}

\onecolumn

\begin{center}
\textbf{\large Supplementary information: Observation of strong backscattering in valley-Hall photonic topological interface modes}
\end{center}

\tableofcontents

\clearpage

\FloatBarrier\section{Structural characterization of bearded valley-Hall waveguides}
This section describes the geometry of the fabricated structures used in the main text and the procedure to characterize them. A similar approach is used to characterize all other waveguides in this work.

\subsection{Overview of samples and waveguide geometry\label{sec:sample-overview}}
\begin{figure}[h]
  \centering
  \includegraphics{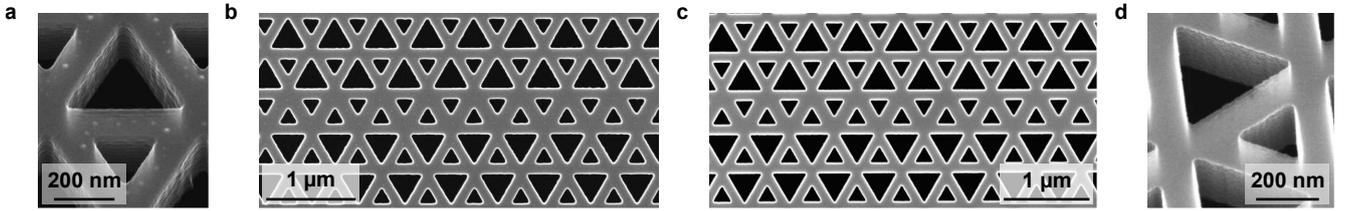}
  \caption{\textbf{Comparision between Sample 1 and 2.} \textbf{a},\textbf{b}, Close-up scanning electron microscope images of the etched triangular holes and top-down view of the interface for Sample 1. \textbf{c},\textbf{d}, Same for Sample 2.\label{fig:comparison-sems}}
\end{figure}

The data underlying the analysis of the main text and here is derived from measurements on two samples, refered to as Sample 1 and Sample 2 (cf. Methods). For comparison, Fig.~\ref{fig:comparison-sems} shows both top-view and tilted scanning electron microscope (SEM) images of bearded valley-Hall (VH) waveguides in Sample 1 (a and b) and Sample 2 (c and d). These images evidence that the improved fabrication process of Sample 2 reduces sidewall roughness and eliminates the bright residues visible on Sample 1 which may result from the electron-beam resist and we expect to have a negligible effect on the optical properties of the waveguides. The electron-beam lithographic mask for both samples use the same nominal crystal geometry with lattice constant $a_0=\SI{512}{nm}$ and hole sidelengths $s_1 = 0.7a_0/\sqrt{3} = 0.40a_0$ and $s_2=1.3/\sqrt{3}=0.75a_0$ (cf. Fig.~2a of the main text). This corresponds to the geometry proposed in Ref.~\cite{Yoshimi_SlowLightWaveguides_2020} but scaled by a factor of $0.965$. It is essential that the band gap opened by the perturbation distinguishing $s_1$ from $s_2$ remains open even in the presence of stochastic fabrication effects. In our case, the random disorder in the transferred pattern is much smaller than the difference between the two hole sizes. Additionally, if the larger triangular hole is made too big, the structural integrity of the device may be compromised. Of the 51 structures fabricated using this design, none suffered structural failure.

\begin{figure}[h]
  \centering
  \includegraphics{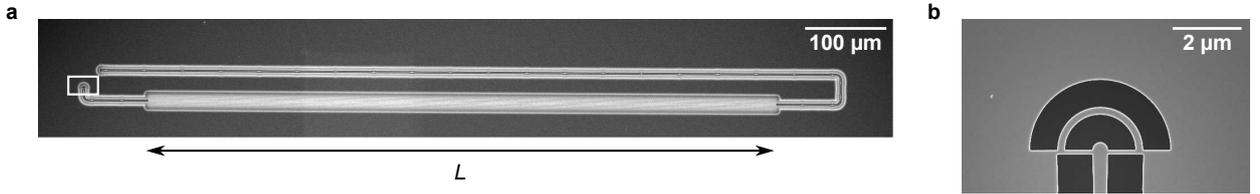}
  \caption{\textbf{Example test circuit.} \textbf{a}, Scanning electron microscope (SEM) image of a test circuit including a photonic topological insulator (PTI) waveguide of length $L=1500a_0$ (indicated on the image) where $a_0$ is the lattice constant.
  \textbf{b}, Top-down SEM image of the circular grating coupler used to couple in and out of the circuit.\label{fig:circuit-overview}}
\end{figure}

\begin{figure}
  \centering
  \includegraphics{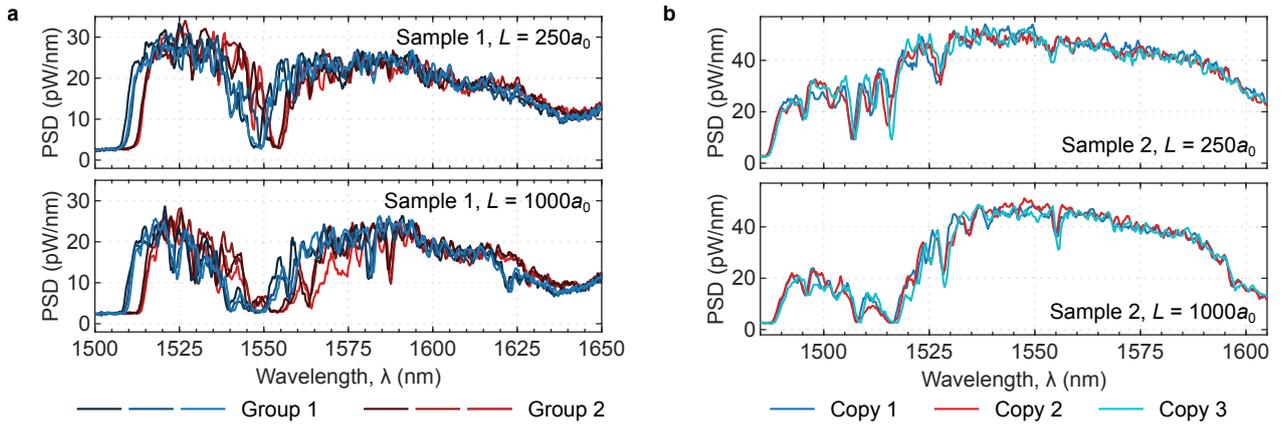}
  \caption{\textbf{Measured raw power spectral densities for nominally identical circuits on Sample 1 and 2.}
  \textbf{a}, Power spectral densities (PSDs) for nominally identical circuits on Sample 1 of length $L=250a_0$ (top) and $L=1000a_0$ (bottom). The two groups are indicated by shades of blue and red, showing the spectral shifts between the two.
  \textbf{b}, Measured PSDs for nominally identical circuits on Sample 2 of length $L=250a_0$ (top) and $L=1000a_0$ (bottom). The colors (blue, red and cyan) denote individual copies.
  \label{fig:spectral-shifts}}
\end{figure}

Both samples contain test circuits which connect to a photonic topological insulator (PTI) waveguide using free-space grating couplers~\cite{Faraon_CoherentGenerationNonclassical_2008} and rectangular strip waveguides, an example of which is shown in Fig.~\ref{fig:circuit-overview}a. A detail of the grating coupler is shown in Fig.~\ref{fig:circuit-overview}b. The circuits are arranged on the chip in arrays, where the length $L$ is swept from $L=250a_0$ to $L=1250a_0$ on Sample 1 and from $L=250a_0$ to $L=1750a_0$ on Sample 2. The lengths of the connecting rectangular waveguides are kept constant to keep their losses constant. Shunt circuits in which the PTI waveguide is fully removed are also fabricated to evaluate coupling efficiency into the PTI waveguide. Sample 1 additionally contains two arrays of test circuits (lengths $L=250a_0$ to $L=1250a_0$) where sharp bends have been introduced near both ends of the PTI waveguide. The raw measured power spectral densities (PSDs) for two lengths of test circuits (with no bends) on both samples are shown in Fig.~\ref{fig:spectral-shifts}. The devices of Sample 1 were fabricated in two groups of three arrays each, spatially separated by about \SI{5}{mm}. The effect of such spatial arrangement is clearly seen in Fig.~\ref{fig:spectral-shifts}a, where we observe systematic spectral shifts of about \SI{6}{nm} between the two groups (indicated by blue and red curves). This is likely a result of either inhomogeneities in device layer thickness or variable loading effects during reactive-ion etching and not of stochastic fabrication disorder. Therefore, we analyze these two groups separately. The test circuits on Sample 2 are fabricated in closer proximity and as evidenced by Fig.~\ref{fig:spectral-shifts}b, exhibit stochastic variations between circuits but no clear systematic spectral shifts.

\FloatBarrier\subsection{Scanning electron microscope characterization and dispersion relations\label{sec:sem-trace}}
\begin{figure}
  \centering
  \includegraphics{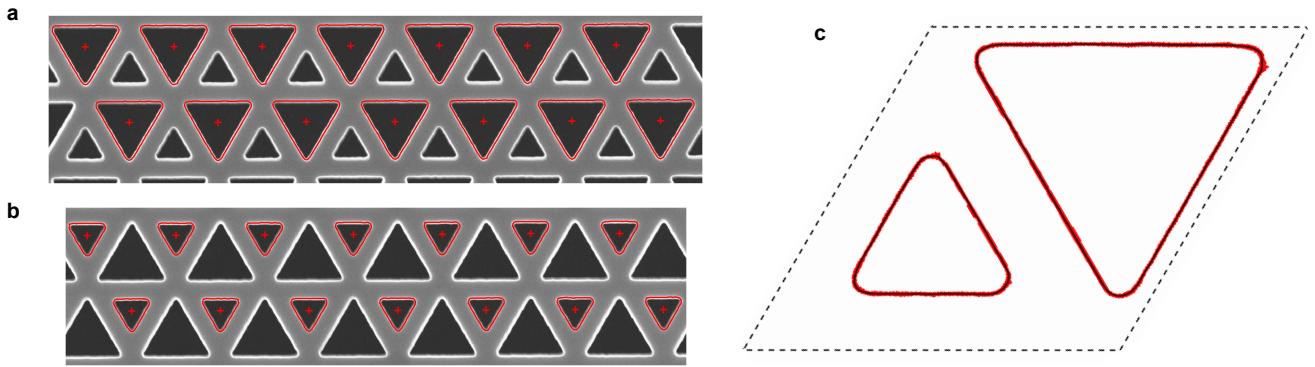}
  \caption{\textbf{Extraction of geometry outline from scanning electron microscope images of Sample 2.}
  \textbf{a},\textbf{b}, Top-down images used to extract a set of large and small triangular hole outlines and centroids. The extracted outlines and centroids are superimposed as red lines and crosses.
  \textbf{c}, Points of outlines of both small and large extracted holes shifted to common centroid and superimposed (red). The extracted average outline is shown (solid black lines) as well as the lattice unit cell (dashed black line). \label{fig:sem-trace-new}}
\end{figure}

\begin{figure}
  \centering
  \includegraphics{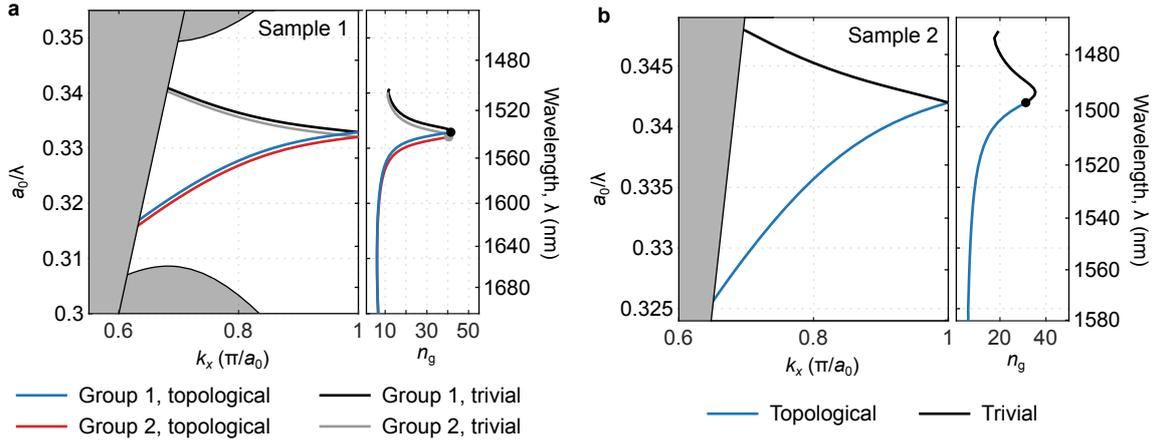}
  \caption{\textbf{Dispersion relations for Sample 1 and 2.}
  \textbf{a}, The dispersion and group index, $n_\mathrm{g}$, for the PTI waveguide on Sample 1 as computed from the outlines extracted by scanning electron microscope (SEM) analysis. Due to the shifts, the dispersions of the two separate groups are computed and plotted separately (see legend). The light cone and bulk bands are indicated by gray shaded regions.
  \textbf{b}, The dispersion and group index, $n_\mathrm{g}$, of waveguides on Sample 2 as computed from the SEM-extracted outlines (shown in Fig.~\ref{fig:sem-trace-new}).
  \label{fig:sem-dispersions}}
\end{figure}

In order to accurately model the dispersion of the topological waveguides under investigation, we extract the geometry of the fabricated samples using SEM. Figure~\ref{fig:sem-trace-new}a and b show top-down SEM images of both Sample 1 and 2. We then analyze the images to extract the outlines and centroids of several holes. Absolute dimensions are fixed by correlating the known lattice constant ($a_0 = \SI{512}{nm}$) with the lattice spanned by the centroids of the extracted features, since we consider this measure to be more accurate than the dimensions obtained by SEM (although we find that the two values do generally agree). To extract the average hole shape, the points from several holes are aggregated by shifting them to a common centroid and averaging points within \SI{3}{degree} angles from the common centroid. This yields averaged hole outlines as displayed in Fig.~\ref{fig:sem-trace-new}c. This procedure is carried out for several SEM images on waveguides located at different positions accross the whole chip. We use the traced hole outlines for the calculation of the band structures for both Samples 1 and 2. We assume vertical sidewalls and a nominal thickness of $h= \SI{220}{nm}$. Figure \ref{fig:sem-dispersions} shows the band structures for both samples. The band diagram for Sample 1 (Fig.~\ref{fig:sem-dispersions}a) includes bands for both groups of circuits (see Supplementary Section~\ref{sec:sample-overview}), showing a 4 nm difference in the degeneracy point, not far from the 6 nm we observe experimentally. For both samples, a small experiment-theory offset of around 6 and 20 nm still persists after the SEM analysis, which is much smaller than the one found when mask parameters are considered (about \SI{110}{nm} for Sample 2). The analysis is therefore necessary since it gives validity to the rigid shift of the band structure we use in Eq.~(3) in the main text.

\begin{figure}[h]
  \centering
  \includegraphics{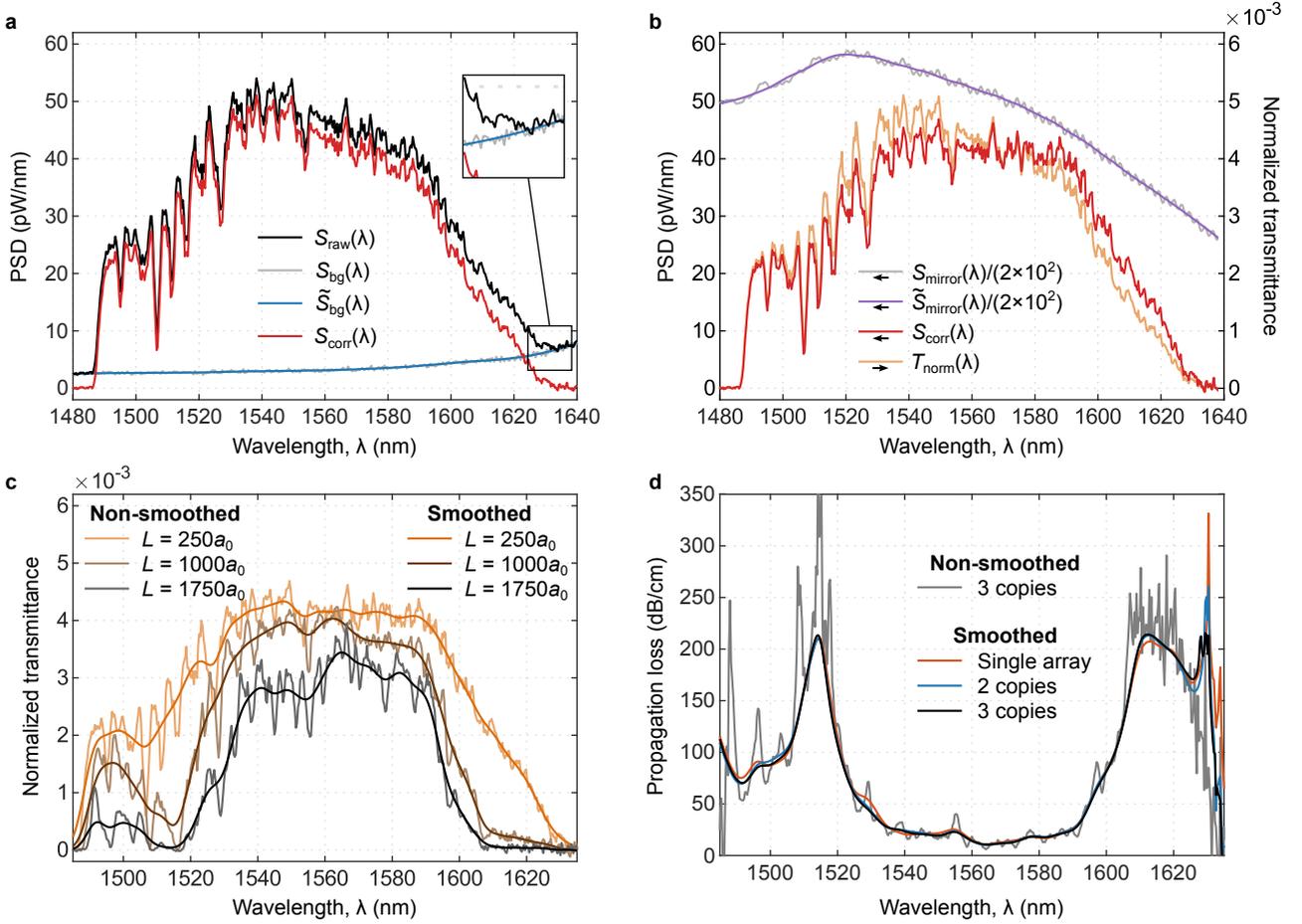}
  \caption{\textbf{Data analysis procedure.}
    \textbf{a}, Raw data of measured power spectral density (PSD), $S_{\mathrm{raw}}(\lambda)$, from a test circuit on Sample 2 with $L=250a_0$ (black). The background PSD, $S_{\mathrm{bg}}(\lambda)$, is shown before (gray) and after (blue) application of the smoothing, i.e., $S_{\mathrm{bg}}(\lambda)$ and $\widetilde{S}_{\mathrm{bg}}(\lambda)$. The background-corrected PSD, $S_{\mathrm{corr}}(\lambda)$, (red) is computed as the difference between the measured PSD and the background PSD.
    \textbf{b}, The background-corrected PSD as obtained from \textbf{a} is shown (red, left $y$-axis), as is the PSD obtained from the silver mirror before (gray) and after (purple) application of the smoothing, i.e., $S_{\mathrm{mirror}}(\lambda)$ and $\widetilde{S}_{\mathrm{mirror}}(\lambda)$. The mirror PSDs are scaled by a factor of $1/(2\times10^{2})$ for display. Their ratio, i.e., the normalized circuit transmittance, $T_{\mathrm{norm}}(\lambda)$, is also shown (beige, right $y$-axis).
    \textbf{c}, The normalized circuit transmittance for three waveguide lengths ($250a_0$ in orange, $1000a_0$ in brown and $1750a_0$ in black) before (lighter curves) and after (darker curves) smoothing. The example smoothed spectra are included alongside other circuit lengths in Fig.~3f of the main text.
    \textbf{d}, The propagation loss as fitted to a restricted data set of 1 (red) or 2 copies (blue) as well as the propagation loss obtained from the full data set of 3 copies (black). Also shown is the propagation loss obtained from the data without the smoothing filter applied to the transmittance data (gray). The black line in \textbf{d} is shown with statistical uncertainty in Fig.~4a of the main text.
    \label{fig:data-treatment}}
\end{figure}

\FloatBarrier\section{Data analysis\label{sec:main-analysis}}
In this section, we will describe the steps of the analysis of the main text and provide additional details. Similar analyses are applied to other waveguide geometries.

\subsection{Background correction and transmittance normalization\label{sec:data-transform}}
The procedure for obtaining the optical spectra is described in Methods. For the measurements on Sample 2, this yields a set of $501$ points describing the PSD, $S_{\mathrm{raw}}(\lambda)$, of the collected light sampled uniformly in wavelength, $\lambda$, from \SI{1480}{nm} to \SI{1640}{nm}. Although light collected at the output path is spatially and polarization filtered, the raw PSD data includes some amount of light scattered from other spatial locations at the chip surface (or elsewhere in the free-space optical setup). Hence, we acquire a PSD spectrum, $S_{\mathrm{bg}}(\lambda)$, from a location on the sample without structures. We assume the minor oscillations in the signal to be uncorrelated with $S_{\mathrm{raw}}(\lambda)$, so we smooth $S_{\mathrm{bg}}(\lambda)$ using a wide ($101$ points $\sim \SI{30}{nm}$) degree-$2$ Savitzky-Golay filter to remove them and obtain the smoothed background $\widetilde{S}_{\mathrm{bg}}(\lambda)$. This is subtracted, isolating the background-corrected PSD as 
\begin{equation}
  \label{eq:1}
  S_{\mathrm{corr}}(\lambda) = S_{\mathrm{raw}}(\lambda) - \widetilde{S}_{\mathrm{bg}}(\lambda)\,.
\end{equation}
The used background spectrum (before and after smoothing) as well as a characteristic circuit spectrum (before and after subtraction of the background) is shown in Fig.~\ref{fig:data-treatment}a. Comparing this background spectrum to the device spectrum outside the transmission band shows good agreement, indicating that this measured spectrum describes the measurement background well. To precisely know the circuit transmittance from input to output grating coupler, we account for the losses and possible dispersion in the optical elements of the measurement setup by acquiring a normalization PSD, $S_{\mathrm{mirror}}(\lambda)$, measured from a silver mirror (cf. Methods), which has an approximately dispersion-free reflectance in the band of interest as well as a weak dependence on the angle of incidence. We again assume the minor oscillations in the signal to be uncorrelated with $S_{\mathrm{raw}}(\lambda)$ and smooth $S_{\mathrm{mirror}}(\lambda)$ with a filter identical to the abovementioned to obtain $\widetilde S_{\mathrm{mirror}}(\lambda)$. From there, the normalized circuit transmittance, $T_{\mathrm{norm}}(\lambda)$, is defined as
\begin{equation}
  \label{eq:3}
  T_{\mathrm{norm}}(\lambda) = \frac{S_{\mathrm{corr}}(\lambda)}{\widetilde{S}_{\mathrm{mirror}}(\lambda)}\,.
\end{equation}
The quantities $S_{\mathrm{corr}}(\lambda)$, $S_{\mathrm{mirror}}(\lambda)$, $\widetilde{S}_{\mathrm{mirror}}(\lambda)$, and $T_{\mathrm{norm}}(\lambda)$ are displayed in Fig.~\ref{fig:data-treatment}b.

The procedure described in the previous paragraph is repeated for a measured PSD for every circuit, reusing the same $S_{\mathrm{bg}}(\lambda)$ and $S_{\mathrm{mirror}}(\lambda)$. The previously discussed example transmittance spectrum, as well as correspondingly treated spectra for PTI waveguides of length $L=1000a_0$ and $L=1750a_0$, are shown in Fig.~\ref{fig:data-treatment}c. Since the spectra exhibit significant fringes likely due to reflections in the circuit, we convolute each transmittance, $T_{\mathrm{norm}}(\lambda)$, with a normalized Gaussian kernel with standard deviation $\sigma=\SI{2.5}{nm}$ to preserve spectral resolution while reducing fringes. We thus obtain the smoothed normalized transmittances, $T(\lambda)$, as the (numerically evaluated) integral
\begin{equation}
  \label{eq:2}
  T(\lambda) = \int d\!\lambda'\, Ne^{-(\lambda-\lambda')^2/2\sigma^2}T_{\mathrm{norm}}(\lambda')\,,
\end{equation}
where the constant $N$ is chosen to normalize the kernel numerically. Figure~\ref{fig:data-treatment}c illustrates the effect of this smoothing filter for three different waveguide lengths. A complete set of smoothed spectra are shown in Fig.~3 of the main text.

\subsection{Extraction of the propagation loss\label{sec:propagation-loss-fit}}
Referring to Fig.~\ref{fig:circuit-overview}a and the normalization procedure described above, we model the transmittance for a specific realization of a circuit of length $L$, $T_{\mathrm{circuit},L}(\lambda)$, as the following product of transmittances:
\begin{equation}
  \label{eq:5}
  T_{\mathrm{circuit},L}(\lambda) = T_{\mathrm{GC,out}}(\lambda)\times T_{\mathrm{WG,out}}(\lambda)\times T_{\mathrm{coupling}}(\lambda)\times T_{\mathrm{prop},L}(\lambda)\times T_{\mathrm{coupling}}(\lambda)\times T_{\mathrm{WG,in}}(\lambda)\times T_{\mathrm{GC,in}}(\lambda)\,,
\end{equation}
where $T_{\mathrm{GC,in}}(\lambda)$ and $T_{\mathrm{GC,out}}(\lambda)$ describe the loss in the input and output free-space couplers, $T_{\mathrm{WG,in}}(\lambda)$ and $T_{\mathrm{WG,out}}(\lambda)$ describe the loss in the strip waveguide before and after the PTI waveguide, $T_{\mathrm{coupling}}(\lambda)$ describe the coupling loss into and out of the PTI waveguide, and $T_{\mathrm{prop},L}(\lambda)$ describes the propagation loss in the PTI waveguide. For simplicity, we have neglected reflections. The transmittances of Eq.~\eqref{eq:5} are all generally stochastic quantities which vary between even nominally identical circuits. We assume them to be independent, allowing us to separate and combine the expectation values of their logarithms freely:
\begin{equation}
  \label{eq:15}
  \begin{split}
    \langle\ln T_{\mathrm{circuit},L}(\lambda)\rangle &= 
    \langle\ln T_{\mathrm{GC,out}}(\lambda)\rangle+ \langle\ln T_{\mathrm{WG,out}}(\lambda)\rangle+ \langle\ln T_{\mathrm{coupling}}(\lambda)\rangle \\
    &\qquad + \langle\ln T_{\mathrm{prop},L}(\lambda)\rangle+ \langle\ln T_{\mathrm{coupling}}(\lambda)\rangle+ \langle\ln T_{\mathrm{WG,in}}(\lambda)\rangle+ \langle\ln T_{\mathrm{GC,in}}(\lambda)\rangle\,.
  \end{split}
\end{equation}
We assume (and observe experimentally) the expectation value of the log-transmittance through the PTI waveguide to depend linearly on $L$, i.e.
\begin{equation}
  \label{eq:7}
  \langle \ln T_{\mathrm{prop},L}(\lambda)\rangle = -\alpha(\lambda)L\,,
\end{equation}
where $\alpha(\lambda)$ is the propagation constant. The respective length and number of all other optical paths and components, e.g., suspension tethers, other than the topological waveguides are kept unchanged in order to rigourously separate transmittance from coupling efficiencies. We can combine remaining transmittances, which are thus independent of $L$, into a parameter, $T_0$, given by
\begin{equation}
  \label{eq:6}
  \ln T_0(\lambda) = \langle\ln T_{\mathrm{GC,out}}(\lambda)\rangle+ \langle\ln T_{\mathrm{WG,out}}(\lambda)\rangle+ \langle\ln T_{\mathrm{coupling}}(\lambda)\rangle
  + \langle\ln T_{\mathrm{coupling}}(\lambda)\rangle+ \langle\ln T_{\mathrm{WG,in}}(\lambda)\rangle+ \langle\ln T_{\mathrm{GC,in}}(\lambda)\rangle\,,
\end{equation}
which we term the constant insertion loss. Hence, we model the ensemble-averaged circuit transmittance as
\begin{equation}
  \label{eq:8}
  \langle\ln T_{\mathrm{circuit},L}(\lambda)\rangle = -\alpha(\lambda)L + \ln T_0(\lambda)\,.
\end{equation}

The analyzed data set consists of measurements of circuit transmittance, $T(\lambda,L)$, as a function of length, $L$, and wavelength, $\lambda$, which are obtained through the filtering procedure described in Supplementary Section~\ref{sec:data-transform}. The data set has three independent datapoints per length-wavelength combination. For each wavelength, we obtain the linear least-squares fit to $\langle\ln T(\lambda,L)\rangle$, assuming it to follow the relation
\begin{equation}
  \label{eq:4}
  \langle\ln T(\lambda,L)\rangle = -\alpha(\lambda)L + \ln T_0(\lambda)\,.
\end{equation}

To illustrate the effect of the number of nominally identical copies, Fig.~\ref{fig:data-treatment}d shows the main fitted propagation length as well as the resulting propagation length where the fitted data has been artificially restricted to only one or two copies of the test circuit array. The similarity between three curves evidences that the extracted propagation loss is only weakly dependent on the number of instances even when these are only a few, implying a narrow probability density function for the waveguide transmission as well as highly reproducible fabrication and optical alignment. The figure also shows the propagation length extracted from the data of the full three copies but without any smoothing applied to the transmittance spectra. The presence of the fringes in the raw data is visible and we see that reported losses in the region of high loss can be considered a lower bound.

\subsection{Evaluation of coupling loss into the photonic topological insulator waveguides\label{sec:insertion-loss}}

\begin{figure}[t]
  \centering
  \includegraphics[width=\textwidth]{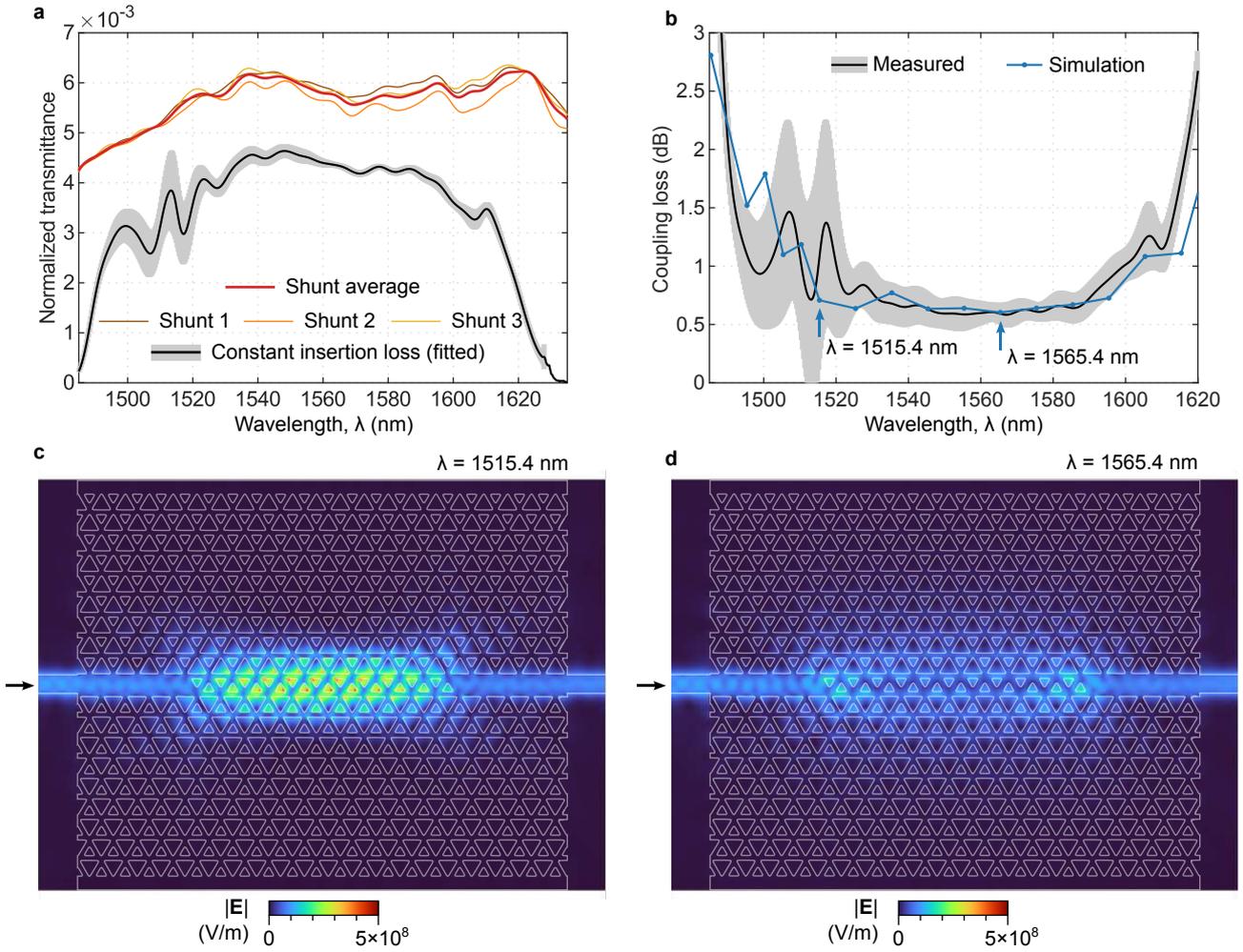}
  \caption{\textbf{Coupling loss into photonic topological insulator waveguides.}
    \textbf{a}, Constant insertion loss, $T_0$, of PTI circuits extracted as the constant term in the fit of the propagation length (black line) with standard error (gray shaded area). The transmittance of three shunt circuits is measured (thin lines, see legend) and the average shunt transmittance (red line) is computed.
    \textbf{b}, Coupling loss (black line) derived from the data in \textbf{a} and coupling loss obtained from numerical simulations (connected blue dots).  Simulated wavelengths have been offset by \SI{15.4}{nm} to align the degeneracy points between simulation and experiment. The points corresponding to the simulations shown in \textbf{c} and \textbf{d} are indicated by arrows. The uncertainty in the measurement is indicated by the gray shading.
    \textbf{c}, Simulated geometry and electric field amplitude at $\lambda =\SI{1515.4}{nm}$ (offset from \SI{1500}{nm}) corresponding to $n_{\mathrm{g}} = 25$.
    \textbf{d}, Simulated geometry and electric field amplitude at $\lambda =\SI{1565.4}{nm}$ (offset from \SI{1550}{nm}) corresponding to $n_{\mathrm{g}} = 7$.
    \textbf{c},\textbf{d}, The displayed field strength is normalized to an injected power of \SI{1}{W}.
    \label{fig:pti-insertion-loss}}
\end{figure}

When fitting the transmittance of the range of PTI waveguide lengths, the constant insertion loss, $T_0$, describes the transmittance of the test circuit with all propagation loss in the PTI waveguide removed and is displayed in Fig.~\ref{fig:pti-insertion-loss}a (solid black line). On the same sample, next to the circuits with PTI waveguides, we include also shunt circuits, equivalent to the test circuits except for the removal of the PTI waveguide and its coupling regions. We measure and transform the transmittance spectra of three nominally identical shunts (displayed in Fig.~\ref{fig:pti-insertion-loss}a) and compute their normalized transmittance spectra, $T^{(\mathrm{shunt})}(\lambda)$, as
\begin{equation}
    T^{(\mathrm{shunt})}(\lambda) = \int d\!\lambda'\, Ne^{-(\lambda-\lambda')^2/2\sigma^2}\left(\frac{S^{(\mathrm{shunt})}_{\mathrm{raw}}(\lambda') - \widetilde{S}_{\mathrm{bg}}(\lambda')}{{\widetilde{S}_{\mathrm{mirror}}(\lambda')}}\right)\,,
\end{equation}
where $S^{(\mathrm{shunt})}_{\mathrm{raw}}(\lambda)$ denotes the measured PSD from a shunt and the rest of the symbols have the same meaning as in Supplementary Section~\ref{sec:data-transform}.
The quantity $T^{(\mathrm{shunt})}(\lambda)$ for each of the three measured shunts as well as their average, $\langle T^{(\mathrm{shunt})}(\lambda)\rangle$, is displayed in Fig.~\ref{fig:pti-insertion-loss}a. By comparison of the fitted constant insertion loss, $T_0(\lambda)$, of a PTI test circuits to the average transmittance of the shunt circuits, we may obtain the coupling efficiency, $\eta_{\mathrm{coupling}}(\lambda)$, as 
\begin{equation}
    \eta_{\mathrm{coupling}}(\lambda) = \sqrt{\frac{T_0(\lambda)}{\langle T^{(\mathrm{shunt})}(\lambda)\rangle}}\,,
\end{equation}
and subsequently the coupling loss as
\begin{equation}
    \langle T_{\mathrm{coupling}}(\lambda) \rangle = 1-\eta_{\mathrm{coupling}}(\lambda)\,.
\end{equation}
The extracted coupling loss from the strip to the PTI waveguide is shown in Fig.~\ref{fig:pti-insertion-loss}b and reach values as low as \SI{0.6}{dB}. The measured coupling loss exhibits some fringes between \SI{1500}{nm} and \SI{1520}{nm} which are likely due to internal reflections. Fig.~\ref{fig:pti-insertion-loss}b also includes the coupling loss $1-\eta_{\mathrm{coupling,sim}}(\lambda)$ calculated using three-dimensional frequency-domain finite-element simulations (see Methods), showing good quantitative agreement. Practically, we model a short section of waveguide ($11.5$ lattice periods) connected to the intermediate coupling waveguides of length $5$ lattice periods (as in the fabricated devices), which are in turn connected to rectangular strip waveguides of width \SI{400}{nm}, and assume the input and output coupling efficiencies to be equal, i.e. $\eta_{\mathrm{coupling,sim}}(\lambda) = \sqrt{T_{\text{sim}}(\lambda)}$. The critical sizes correspond to those in the fabricated structure. The numerical coupling loss are offset in wavelength by \SI{15.4}{nm} to align the degeneracy point of the simulated structure with that found in the experiment (see Supplementary Section~\ref{sec:sem-trace}). Fig.~\ref{fig:pti-insertion-loss}c and d show the resulting simulated electric field strength for the two wavelengths highlighted with arrows in Fig.~\ref{fig:pti-insertion-loss}b. We note that the spectral shift is different from the one obtained for the propagation loss. This is likely due to perturbations of the transmittance by geometrical variations and/or the surface oxide, which is expected to be dispersive due to the significant differences in the spatial structure of the modes at different wavelengths. 

\FloatBarrier\section{Propagation loss measurements on other waveguides}
The main propagation loss analysis is performed on Sample 2 and on waveguides using the bearded valley-Hall (VH) interface first proposed in Ref.~\cite{Yoshimi_SlowLightWaveguides_2020}. To provide a link between the waveguides on Sample 1 and Sample 2, we repeat the analysis of the main text for the nominally identical topological waveguides on Sample 1, which have slightly different dispersion (see Supplementary Section \ref{sec:sem-trace}). To compare the quality of our fabricated samples to previous experiments, we measure the propagation loss on standard conventional W1 waveguides and demonstrate that our process is state of the art, with the lowest reported propagation losses in the literature being achieved in W1 waveguides on Sample 2.  Finally, we also carry out identical experiments on two additional VH waveguide geometries, including a zigzag interface, for comparison. We find that the conventional W1 waveguides, the standard geometry of which has not been optimized, outperform any measured topological waveguide.

\FloatBarrier\subsection{Experiments on bearded valley-Hall waveguides on Sample 1\label{sec:sample1-analysis}}
\begin{figure}
  \centering
  \includegraphics{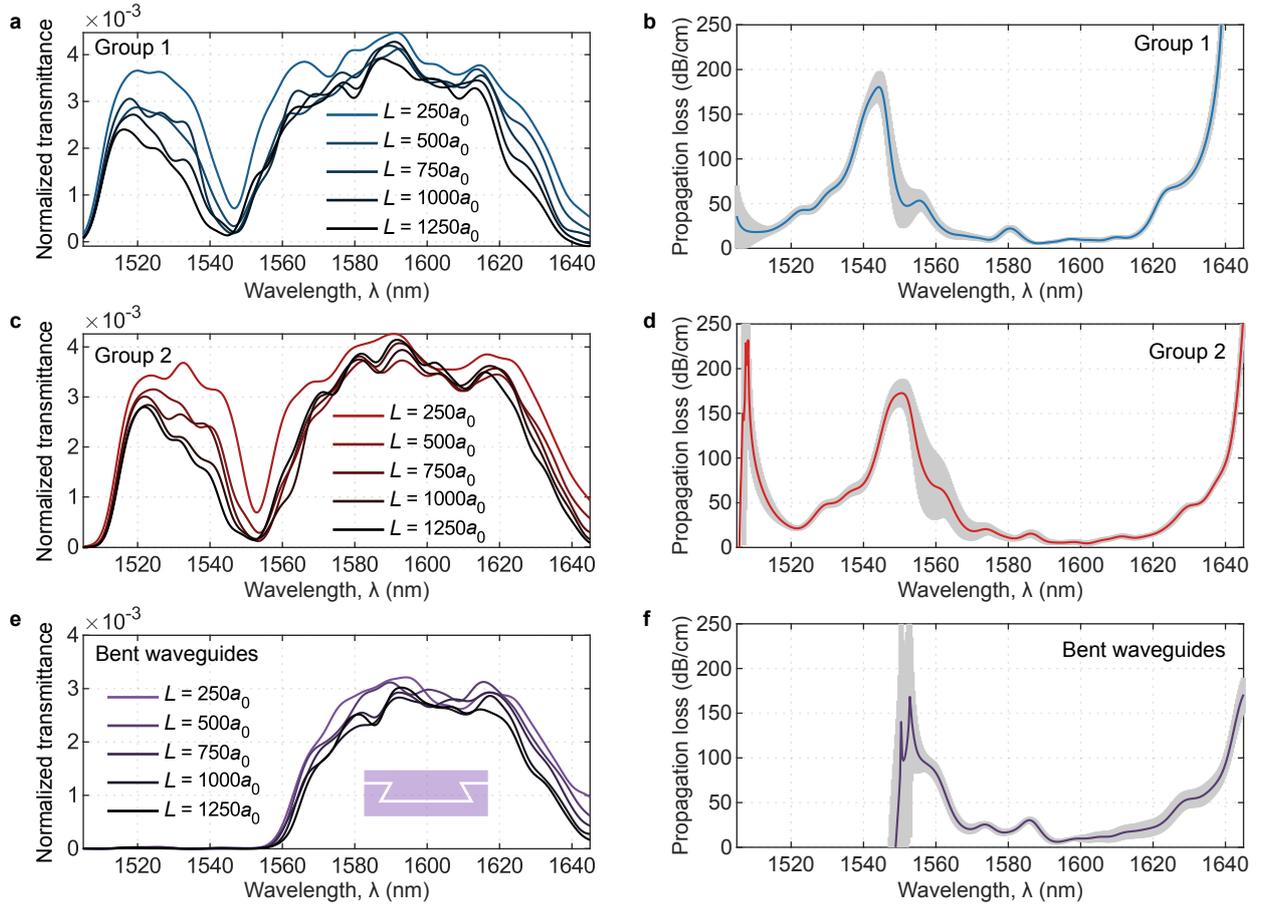}
  \caption{\textbf{Propagation losses of photonic topological insulator waveguides on Sample 1.}
  \textbf{a},\textbf{c}, Measured normalized test circuit transmittance spectra for Group 1 and 2 circuits.
  \textbf{b},\textbf{d}, Propagation loss as fitted from measurements performed on Group 1 and 2, respectively.
  \textbf{e}, Measured normalized transmittance spectra for test circuits which include sharp bends.
  \textbf{f}, Propagation loss as fitted from the transmittance measurements on test circuits that include sharp bends in the photonic topological insulator waveguide.
  \label{fig:sample1-analysis}
  }
\end{figure}

The measurements of propagation losses shown in Figs.~2 and 3 of the main text are performed on Sample 2. To establish that the topological waveguides on both samples operate in essentially the same way, we apply the analysis described in Supplementary Section~\ref{sec:main-analysis} to equivalent transmission measurements on the circuits on Sample 1 to obtain the propagation loss of these structures. Due to the systematic spectral shifts observed (see Supplementary Section~\ref{fig:spectral-shifts}), we perform independent analyses of the measurements for the two spatially offset groups of test circuits. The normalized circuit transmittances (obtained as described in Supplementary Section~\ref{sec:data-transform}) for a single array in Group 1 and 2 are displayed in Fig.~\ref{fig:sample1-analysis}a and c, respectively. The transmittance spectra clearly exhibit two intervals of high transmittance separated by a region of low transmittance and low coupling efficiency, which we indentify as the transition between topological and trivial bands. The fitted propagation length (Fig.~\ref{fig:sample1-analysis}b and d) exhibits more oscillations compared to Sample 2, which is likely due to the more limited span in waveguide lengths on Sample 1 (the length $L$ is varied from $250a_0$ to $1250a_0$ compared to the variation from $250a_0$ to $1750a_0$ for Sample 2) exacerbating the effect of internal circuit reflections on the analysis. We also explore the propagation losses as measured from test circuits where two sets of two sharp turns near the beginning and end of the PTI waveguides are included. The central segments of PTI waveguide between the bends are varied in length as the straight waveguides, allowing for the same analysis to be performed on the topological band for these waveguides. The normalized transmittance spectra of a single array of bent waveguides waveguides is shown in Fig.~\ref{fig:sample1-analysis}e. We apply the equivalent analysis to a total of two arrays of test circuits to obtain the propagation loss shown in Fig.~\ref{fig:sample1-analysis}f, which is in very good quantitative agreement with the data in the straight waveguides for the topological mode.

\FloatBarrier\subsection{Reference measurements on W1 waveguides\label{seq:w1-losses}}
\begin{figure}
  \centering
  \includegraphics[width=\textwidth]{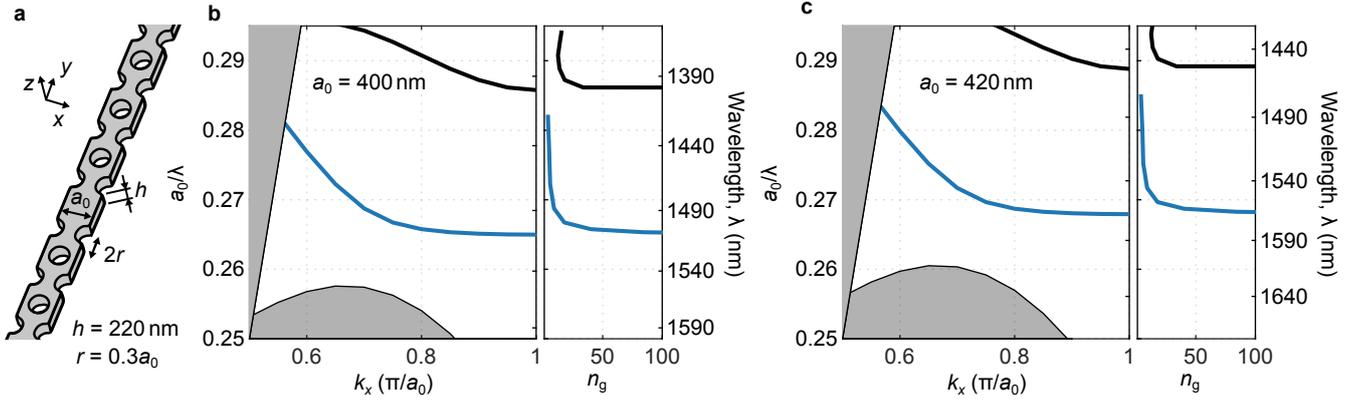}
  \caption{\textbf{Design and dispersion of W1 waveguides.}
    \textbf{a}, Schematic showing the geometry of a single periodic unit of the W1 waveguides used here. The geometrical parameters (lattice constant $a_0$, hole radius $r$, and membrane thickness $h$) are indicated.
    \textbf{b}, Dispersion (left) and group index (right) computed for the supercell with $a_0=\SI{400}{nm}$, $h=\SI{220}{nm}$, and $r=0.3a_0$.
    \textbf{c}, Dispersion (left) and group index (right) computed for the supercell with $a_0=\SI{420}{nm}$, $h=\SI{220}{nm}$, and $r=0.3a_0$.
    \label{fig:w1-dispersion}}
\end{figure}

\begin{figure}
  \centering
  \includegraphics[width=0.9\textwidth]{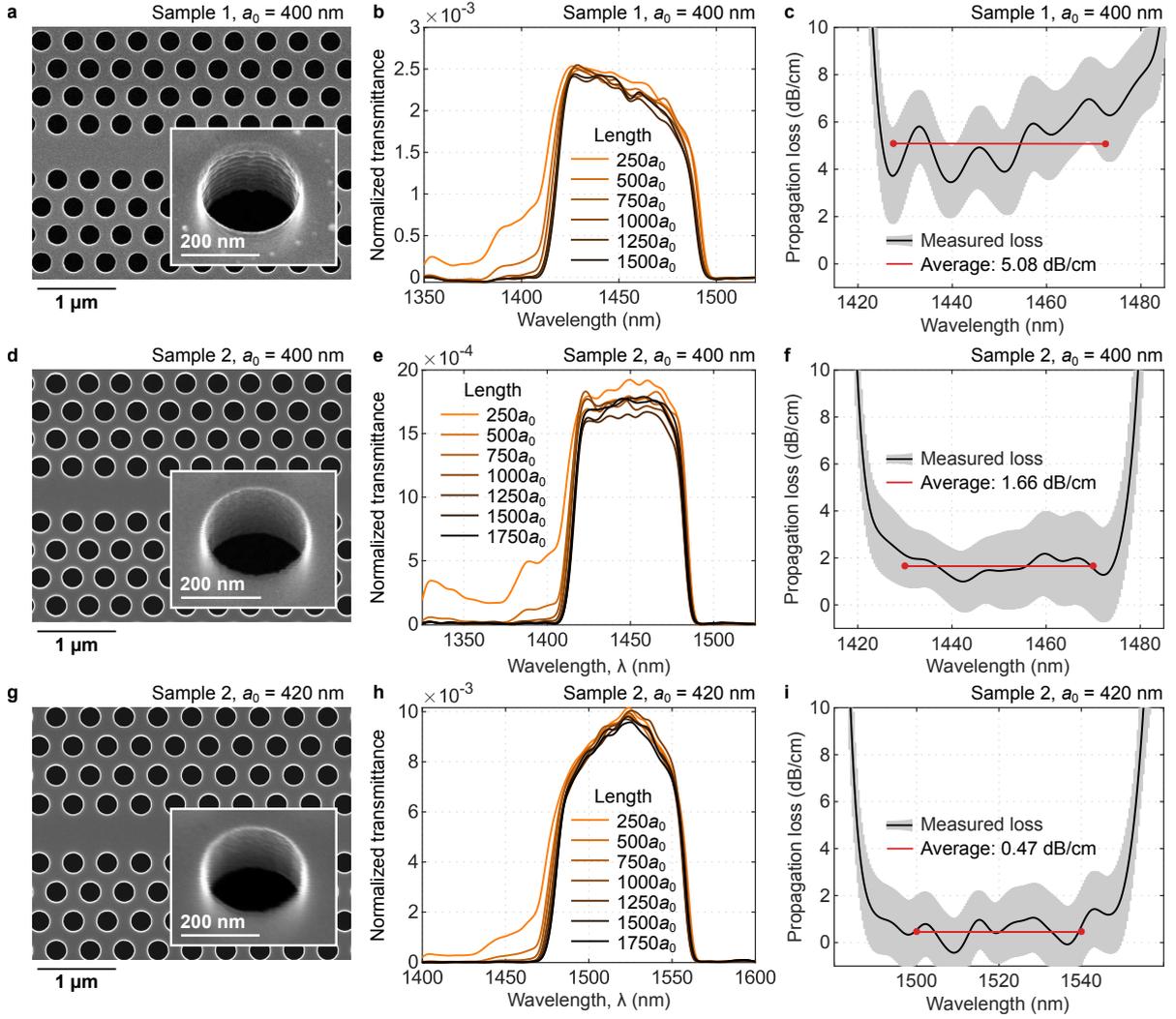}
  \caption{\textbf{Benchmark measurements on W1 waveguides.}
    \textbf{a}, Scanning electron microscope (SEM) image of the W1 waveguide on Sample 1 with the inset showing the details of a single hole.
    \textbf{b}, Transmission measurements of a single array of W1 waveguides on Sample 1, with lengths from $250a_0$ to $1500a_0$ (light to darker, see legend).
    \textbf{c}, Wavelength-dependent propagation loss for W1 waveguides on Sample 1, fitted from the decrease in transmission as a function of length obtained from measurements on three sets of waveguides like the one shown in \textbf{b}. The gray shaded area indicates the standard error on the fits.  The average loss in the interval from \SI{1430}{nm} to \SI{1470}{nm}, \SI{5.08+-0.09}{dB/cm}, is shown with a red line.
    \textbf{d},\textbf{g}, SEM images of the waveguides on Sample 2 with lattice constants $a_0=\SI{400}{nm}$ and $a_0 = \SI{420}{nm}$ respectively. The insets show single holes in the respective devices.
    \textbf{e},\textbf{h}, Normalized measured transmittance spectra for a single array of test circuits on Sample 2 with W1 waveguides with lattice constants $a_0=\SI{400}{nm}$ and $a_0=\SI{420}{nm}$, respectively. The measured circuit lengths span from $L=250a_0$ to $L=1750a_0$ (see legend). 
    \textbf{f}, The measured propagation loss obtained from measurements of three arrays of W1 waveguide test circuits with lattice constants $a_0=\SI{400}{nm}$. The gray shaded area indicates the standard error on the fits. The average loss in the interval from \SI{1430}{nm} to \SI{1470}{nm}, \SI{1.66+-0.03}{dB/cm}, is shown with a red line.
    \textbf{i}, The measured propagation loss obtained from measurements of three arrays of W1 waveguide test circuits with lattice constants $a_0=\SI{420}{nm}$. The gray shaded area indicates the standard error on the fits. The average loss in the interval from \SI{1500}{nm} to \SI{1540}{nm}, \SI{0.47+-0.04}{dB/cm}, is shown with a red line.
    \label{fig:w1-loss-new}}
\end{figure}

Structural defects are extremely hard or even impossible to quantify in practice because the structural and chemical composition would need to be mapped with atomic resolution across hundreds of microns of waveguide. And even if possible, such a procedure has never been attempted and such a measurement would therefore not enable a direct comparison with the levels of disorder present in previously published works in nanophotonics. We therefore benchmark the structural disorder in our samples indirectly by measuring the losses in conventional W1 photonic-crystal waveguides~\cite{MurendranathPatil_ObservationSlowLight_2022} fabricated on the same chips and batches as for the PTI devices. On both samples, we include a set of W1 waveguides formed by removing a row of holes in a triangular lattice with lattice constant $a_0 = \SI{400}{nm}$ and hole radius $r=0.3a_0$. On Sample 2, we additionally include a W1 waveguide of similar design but scaled to $a_0 = \SI{420}{nm}$. Since the structures are fabricated next to the PTI structures explored here, they all have the same nominally \SI{220}{nm} thick device layer a distance of \SI{2}{\micro{}m} from the handle layer. The dispersion diagrams for the nominal structures (i.e., with the dimensions of the mask) are shown in Fig.~\ref{fig:w1-dispersion}b and c, which also show the group index. We fabricate three identical arrays of test circuits (of identical configuration to the PTI circuits as exemplified in Fig.~\ref{fig:circuit-overview}) where the PTI waveguide has been substituted with W1-waveguides of lengths $250a_0$, $500a_0$, $750a_0$, $1000a_0$, $1250a_0$, and $1500a_0$. The arrays on Sample 2 additionally include waveguides of length $1750a_0$. The photonic-crystal waveguides are coupled to suspended strip waveguides using an intermediate region where the lattice spacing has been stretched by a factor of $1.07$ \cite{hugoninCouplingSlowmodePhotonic2007}. Scanning electron microscope images of the three fabricated structures are included in Fig.~\ref{fig:w1-loss-new}a,~d,~and~g, including zoom-ins of a single etched circular hole. The same conclusions about the fabrication process as those drawn from Fig.~\ref{fig:comparison-sems} can be drawn from the W1 waveguides.

We characterize the propagation loss of the W1 waveguides using the same procedure as for the PTI waveguides. The transmittance spectra for a single array of devices of each kind are shown in Fig.~\ref{fig:w1-loss-new}b, e, and h. All spectra evidence a transmission band of approximately \SI{70}{nm}. When the wavelength approaches the band edge, the group index diverges in theory (see Fig.~\ref{fig:w1-dispersion}b and d), which is not reached in experiment because the backscattering also scales with the group index and the system enters the regime of Anderson localization \cite{Topolancik_ExperimentalObservationStrong_2007, Sapienza_CavityQuantumElectrodynamics_2010,Garcia_TwoMechanismsDisorderinduced_2017}. This causes a large increase in propagation loss for all waveguides. On the other side of the transmission band (toward lower wavelengths), the dispersion crosses the light line and the waveguide mode becomes leaky, which is also visible as a large increase in propagation loss. For each wavelength in the spectra of Fig.~\ref{fig:w1-loss-new}b,~e,~and~h, we perform a linear regression to extract the propagation loss, yielding the wavelength-dependent propagation loss shown in Fig.~\ref{fig:w1-loss-new}c,~f,~and~i, respectively. For the waveguides with $a_0=\SI{420}{nm}$ (Sample 2), we observe losses below \SI{1}{dB/cm} in the linear part of the dispersion, i.e., when spectrally far from the slow-light regime. The losses are slightly higher for the waveguides with $a_0=\SI{400}{nm}$ (Sample 2) and still larger for the W1 waveguides in Sample 1. While the first may be due to either the slight differences in group index in the linear regime or different field intensities at the hole boundaries, the higher losses for Sample 1 are very likely due to the differing fabrication processes, which is consistent with the higher level of sidewall roughness directly visible in the SEM (see Fig.~\ref{fig:w1-loss-new}a, d and g). 

For the two W1 waveguides on Sample 2, the measured propagation loss is comparable to or smaller than the statistical error. To obtain a better estimate of the propagation loss, we average the transmission over a \SI{40}{nm} bandwidth across the center of the transmission band. The error is estimated as the standard error on the mean from the variance in the interval. For the waveguides with $a_0=\SI{400}{nm}$ we choose the interval from \SI{1430}{nm} to \SI{1470}{nm} and obtain \SI{5.08+-0.09}{dB/cm} and \SI{1.66+-0.03}{dB/cm} for the waveguides on Sample 1 and Sample 2 respectively. For the waveguide with $a_0=\SI{420}{nm}$, we choose the interval from \SI{1500}{nm} to \SI{1540}{nm} and obtain an average transmission of \SI{0.47+-0.04}{dB/cm}. These averages have been superimposed onto the measured propagation losses in Fig.~\ref{fig:w1-loss-new}c, f, and i.
For W1 suspended silicon waveguides, the previously reported state-of-the-art propagation losses in that region and at wavelengths around \SI{1550}{nm} include \SI{24}{dB/cm} \cite{McNab_UltralowLossPhotonic_2003a}, \SI{5}{dB/cm} \cite{Kuramochi_DisorderinducedScatteringLoss_2005}, \SI{4.1}{dB/cm} \cite{OFaolain_LowlossPropagationPhotonic_2006}, and \SI{2}{dB/cm} \cite{Kuramochi_ScatteringLossPhotonic_2005,Notomi_NonlinearAdiabaticControl_2007}. This means that, despite the differences observed for the losses in the different W1 waveguides, they all exhibit state-of-the-art loss values and by extension, since the W1 waveguides are fabricated alongside the PTI waveguides, that the conclusions drawn from the measurements on the VH PTI waveguides hold for state-of-the-art silicon nanofabrication.

\FloatBarrier\subsection{Experiments on other photonic topological interfaces}
\begin{figure}
  \centering
  \includegraphics{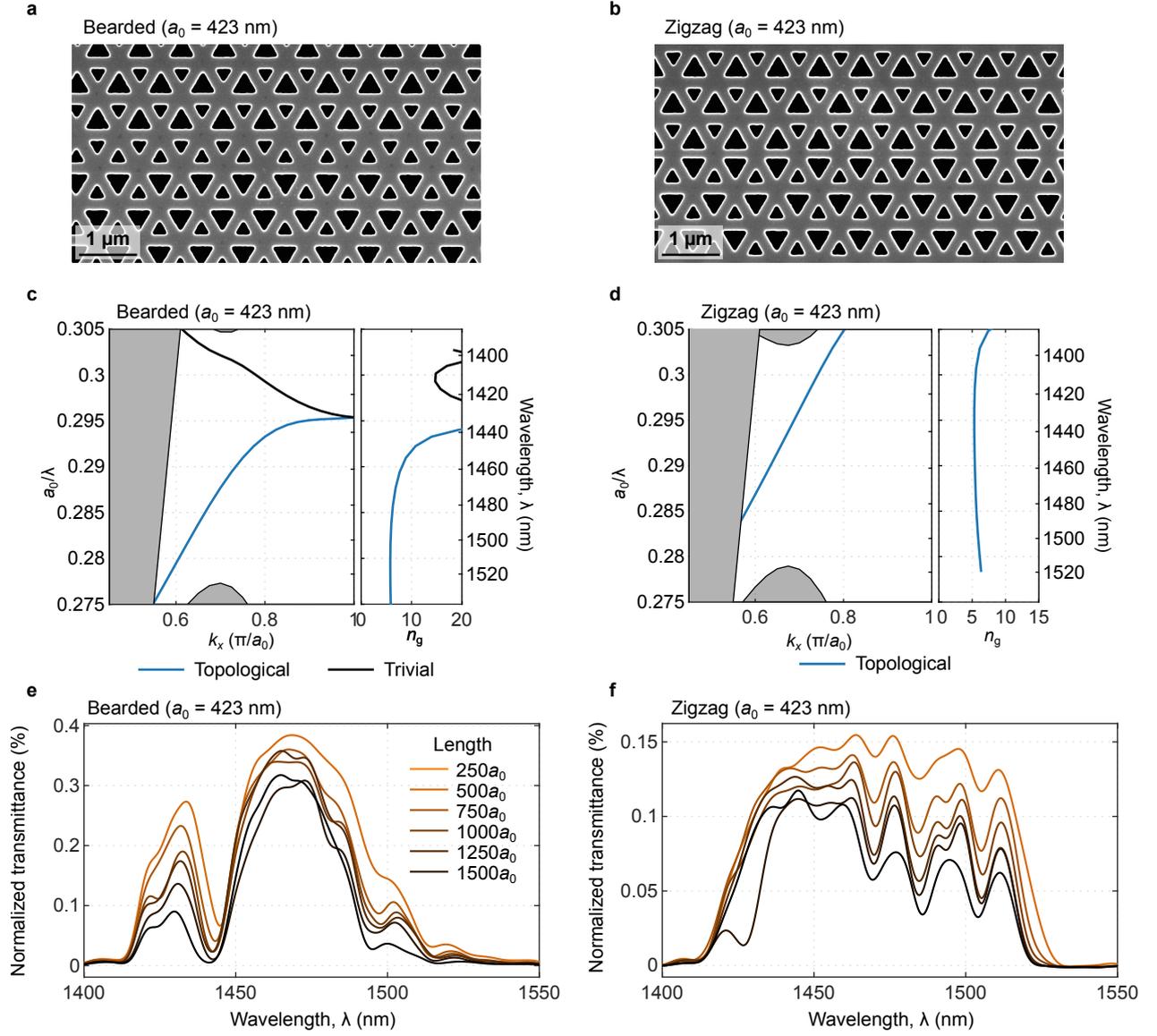}
  \caption{\textbf{Comparison of different valley-Hall interfaces sharing their unit cell.}
    \textbf{a}, Scanning electron microscope (SEM) image of the interface geometry of the measured bearded interface.
    \textbf{b}, Scanning electron microscope image of the interface geometry of the measured zigzag interface.
    \textbf{c}, Dispersion diagram for the bearded interface obtained from the geometry extracted from the SEM image of \textbf{a}.
    \textbf{d}, Dispersion diagram for the zigzag interface obtained from the geometry extracted from the SEM image of \textbf{a}.
    \textbf{e}, Subset of measured transmittance spectra for the measured bearded interface.
    \textbf{f}, Subset of measured transmittance spectra for the measured zigzag interface.
    \label{fig:s19-dispersion}
  }
\end{figure}

\begin{figure}
  \centering
  \includegraphics{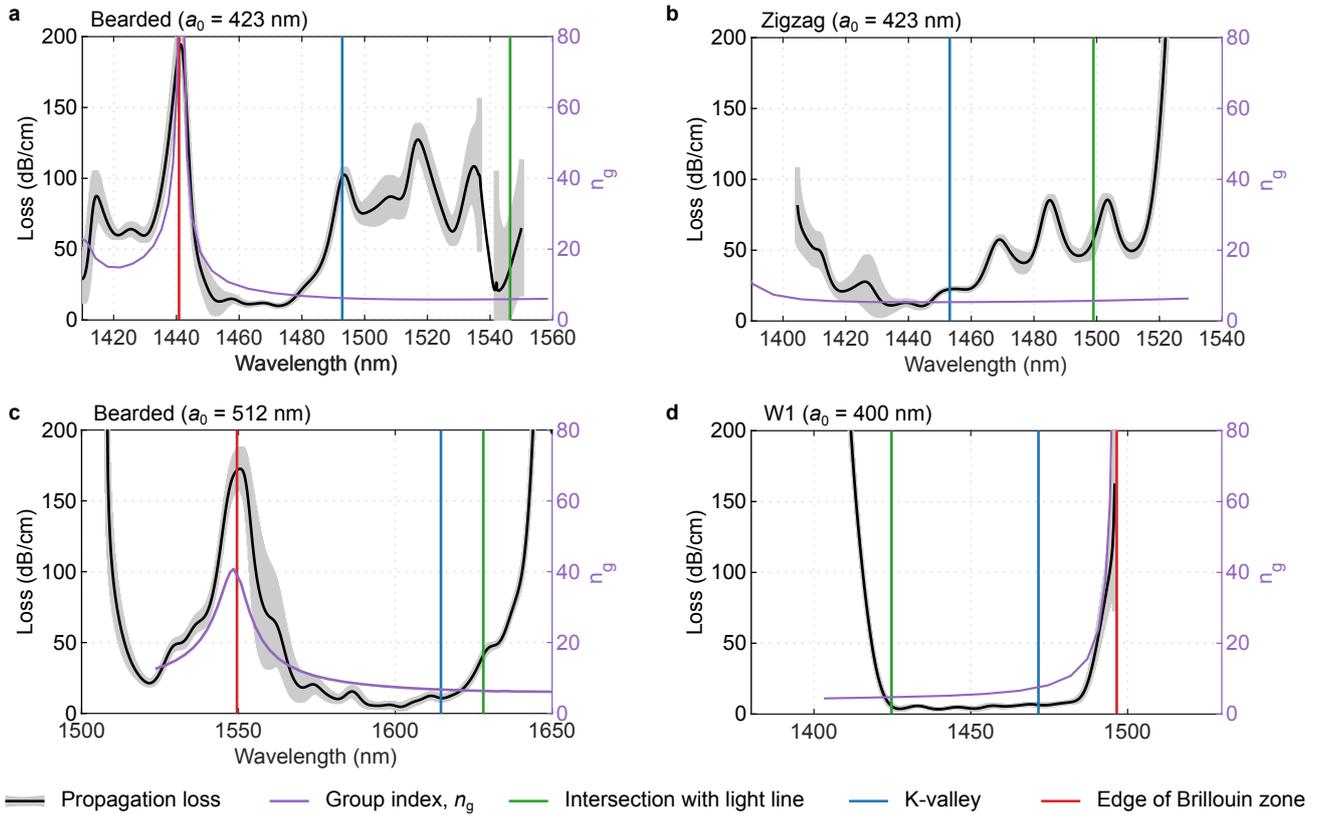}
  \caption{\textbf{Measured propagation loss of topological and conventional waveguides on Sample 1.}
    \textbf{a}, Measured propagation loss (black) and standard deviation (gray) for a W1 waveguide with the calculated group index (purple), the intersection with the light-line (green), the crossing with the edge of the Brillouin zone (blue) and the K-valley (red).
    \textbf{b}, Measured propagation loss (black) and standard deviation (gray) for a zigzag valley-Hall waveguide with the calculated group index (purple), the intersection with the light-line (green) and the K-valley (red). The edge of the Brillouin zone (blue) is not visible in the shown wavelength range.
    \textbf{c}, Measured propagation loss (black) and standard deviation (gray) for a bearded valley-Hall waveguide corresponding to the one used in the main analysis with the calculated group index (purple), the intersection with the light-line (green), the crossing with the edge of the Brillouin zone (blue) and the K-valley (red).
    \textbf{d}, Measured propagation loss (black) and standard deviation (gray) for a bearded valley-Hall waveguide form using the unit cell of \textbf{b} with the calculated group index (purple), the intersection with the light-line (green), the crossing with the edge of the Brillouin zone (blue) and the K-valley (red).
  \label{fig:s19-loss}
  }
\end{figure}

The data presented in the main text was obtained with a device with the bearded interface proposed in Ref.~\cite{Yoshimi_SlowLightWaveguides_2020}. However, several other interfaces, i.e., ways to construct waveguides from juxtaposition of PTIs, exist. A commonly employed interface is the zigzag interface, which, as opposed to the bearded interface, only supports slow light outside the bulk band gap and does not sustain an additional trivial mode with which to compare. It is therefore a less suited geometry to address the existence of topological protection against backscattering. Nevertheless, we provide in this work for completeness additional measurements of propagation losses in both zigzag and bearded interfaces formed from another unit cell, namely that of Ref.~\cite{shalaev_robust_2019}. We retain from Ref.~\cite{shalaev_robust_2019} the lattice constant, $a_0=\SI{423}{nm}$, and the values of $s_1=0.4$ and $s_2=0.6$. These structures were all fabricated in parallel as part of Sample 1, ensuring that they both have equivalent levels of roughness. Figures~\ref{fig:s19-dispersion}a and b show the SEM images of the fabricated waveguide. The dispersion diagrams for these two topological waveguides are calculated based on the SEM images as done for the structure discussed in the main text and described in Supplementary Section~\ref{sec:sem-trace} and are shown in Fig.~\ref{fig:s19-dispersion}c and d. Example measured transmission spectra for various lengths made from the two interfaces are shown in Fig.~\ref{fig:s19-dispersion}e and f. Repeating again the analysis described in Supplementary Section~\ref{sec:main-analysis} yields the propagation losses shown in Fig.~\ref{fig:s19-loss}a and b.

For comparison, the measured propagation losses of other waveguides on Sample 1, namely the bearded interface with $a_0=\SI{512}{nm}$ and the conventional W1 waveguides are shown in Fig.~\ref{fig:s19-loss}c and d. All curves include for reference the group index of the relevant mode and the wavelengths at which the mode crosses the light line, at which its propagation constant corresponds to the projected K-valley, and at the edge of the Brillouin zone. As in the main text, these have been found from the simulated band structures after rigidly shifting them by a fixed number of nanometers. The spectral shift is determined by a fit of the loss peak for the bearded interface. We observe no indication that the propagation loss is minimal at the K-valley in any of the three topological waveguides.

\subsection{Summary of measurements presented in this work}
In total, we present in this work measurements on 9 different waveguide ensembles of both bearded and zigzag topological waveguides as well as standard conventional W1 waveguides. Table~\ref{table:zigzag-measurements-overview} shows an overview of these measurements, including references to relevant dispersion diagrams, transmission spectra, and measured propagation lengths.

For each ensemble (except the W1 waveguides on Sample 2), we locate the point of minimal propagation loss and display the resulting minimal propagation loss in the table. Since the low propagation loss in the W1 waveguides on Sample 2 causes a large relative statistical error in the measurement, we list instead the conservative average propagation loss over a \SI{40}{nm} at the center of the transmission band. For this reason, the listed minimum propagation loss of the W1 waveguides will tend to be higher than the corresponding listed minimum propagation loss for other waveguides. Even so, we find the W1 waveguides to significantly outperform all topological waveguides measured -- both in minimum propagation loss and in bandwidth over which this low propagation loss is sustained. Furthermore, measurements on zigzag and bearded interface waveguides formed from the same unit cell (waveguides with $a_0=\SI{423}{nm}$) are found to have comparable propagation losses. Referring to the graphs of propagation losses (Fig.~\ref{fig:s19-loss} and Figures referred to in Table~\ref{table:zigzag-measurements-overview}), we also observe that the conventional W1 waveguides exhibit less dispersion in the propagation loss than the topological waveguides.

\begin{table}[p]
  \newcolumntype{?}{!{\vrule width 2pt}}
  \newcommand{\mcrot}[4]{\multicolumn{#1}{#2}{\rlap{\rotatebox{#3}{#4}~}}}
  \begin{minipage}{1\textwidth}\renewcommand*\footnoterule{}
    \centering\small
    \begin{tabular}{|c|c|c|c|c||c|c||c|c|c|c|c|}
      \hline
      \rotatebox{90}{Sample}
      & \rotatebox{90}{Topological / conventional\;}
      & \rotatebox{90}{Waveguide geometry}
      & \rotatebox{90}{Lattice constant, $a_0$ (nm)}
      & \rotatebox{90}{\makecell[l]{Ensemble size\\(no. of waveguides)}}
      & \rotatebox{90}{\makecell[l]{Mininimum measured\\propagation loss (dB/cm)}}
      & \rotatebox{90}{\makecell[l]{Wavelength of minimum\\measured propagation\\loss (nm)}}
      & \rotatebox{90}{\makecell[l]{Images of structure\\(fig. no.)}}
      & \rotatebox{90}{\makecell[l]{Dispersion diagram\\(fig. no.)}}
      & \rotatebox{90}{\makecell[l]{Measured transmittance\\(fig. no.)}}
      & \rotatebox{90}{\makecell[l]{Propagation loss\\(fig. no.)}}
      & \rotatebox{90}{\makecell[l]{Out-of-plane scattering\\(fig. no.)}}
      \\
      \hline      
      1 & Topo. & Bearded\footnote{Devices from Group 1 (see Supplementary Section \ref{sec:sample-overview}).} & \num{512}
      & 15
      & \num{6 +- 2}
      & 1588.5
      & 1d\footnote{Scanning electron microscope images of structures from Group 1 and 2 are indistinguishable to the naked eye.\label{fn:eye}}, \ref{fig:comparison-sems}a--b\footref{fn:eye}
      & \ref{fig:sem-dispersions}a
      & 1f, \ref{fig:spectral-shifts}a, \ref{fig:sample1-analysis}a
      & \ref{fig:sample1-analysis}b
      & --
      \\
      1 & Topo. & Bearded\footnote{Devices from Group 2 (see Supplementary Section \ref{sec:sample-overview}).} & \num{512}
      & 15
      & \num{5 +- 3}
      & 1601.7
      & 1d\footref{fn:eye}, \ref{fig:comparison-sems}a--b\footref{fn:eye}
      & \ref{fig:sem-dispersions}a
      & 1f, \ref{fig:spectral-shifts}a, \ref{fig:sample1-analysis}c
      & \ref{fig:sample1-analysis}d, \ref{fig:s19-loss}c
      & --
      \\
      1 & Topo. & Bearded\footnote{Devices with waveguides containing four sharp bends.} & \num{512}
      & 10\footnote{An additional waveguide with four bends was measured for the spectrum in Fig.~1e.}
      & \num{6 +- 3}
      & 1593.6
      & 1c
      & \ref{fig:sem-dispersions}a
      & 1e,
        \ref{fig:sample1-analysis}e
      & \ref{fig:sample1-analysis}f
      & 5
      \\
      1 & Topo. & Bearded & \num{423}
      & 18
      & \num{10 +- 2}
      & 1472.6
      & \ref{fig:s19-dispersion}a
      & \ref{fig:s19-dispersion}c
      & \ref{fig:s19-dispersion}e
      & \ref{fig:s19-loss}a
      & --
      \\
      1 & Topo. & Zigzag & \num{423}
      & 18
      & \num{11 +- 3}
      & 1444.2
      & \ref{fig:s19-dispersion}b
      & \ref{fig:s19-dispersion}d
      & \ref{fig:s19-dispersion}f
      & \ref{fig:s19-loss}b
      & --
      \\
      1 & Conv. & W1 & \num{400}
      & 18
      & \num{4 +- 2}
      & 1439.7
      & \ref{fig:w1-loss-new}a
      & \ref{fig:w1-dispersion}b\footnote{\label{fn:w1-disp}Since we consider the W1 waveguides mainly in the regime of low dispersion, the dispersion diagrams in Fig.~\ref{fig:w1-dispersion} are computed from the nominal geometry.}
      & \ref{fig:w1-loss-new}b
      & \ref{fig:w1-loss-new}c, \ref{fig:s19-loss}d
      & --
      \\
      \hline
      2 & Topo. & Bearded & \num{512}
      & 21
      & \num{11 +- 1}
      & 1439.7
      & 3a--e, $(\ldots)$\footnote{Images of the bearded-interface on Sample 2: Figures 3a--e, \ref{fig:circuit-overview}, \ref{fig:comparison-sems}c--d, \ref{fig:sem-trace-new}, \ref{fig:far-field-straight}a.}
      & 2d, \ref{fig:sem-dispersions}b
      & 3f, \ref{fig:spectral-shifts}a, \ref{fig:data-treatment}a--c
      & 4, \ref{fig:data-treatment}d
      & \ref{fig:far-field-straight}
      \\
      2 & Conv. & W1 & \num{400}
      & 24
      & \num{1.66+-0.03}\footnote{\label{fn:w1-averaging}The minimum measured propagation loss for the W1 waveguides on Sample 2 shows a conservative estimate based on the average transmission in a \SI{40}{nm} band rather than a single point (see Supplementary Section \ref{seq:w1-losses}).}
      & N/A\footref{fn:w1-averaging}
      & \ref{fig:w1-loss-new}d
      & \ref{fig:w1-dispersion}b\footref{fn:w1-disp}
      & \ref{fig:w1-loss-new}e
      & \ref{fig:w1-loss-new}f
      & --
      \\
      2 & Conv. & W1 & \num{420}
      & 24
      & \num{0.47+-0.04}\footref{fn:w1-averaging}
      & N/A\footref{fn:w1-averaging}
      & \ref{fig:w1-loss-new}g
      & \ref{fig:w1-dispersion}c\footref{fn:w1-disp}
      & \ref{fig:w1-loss-new}g
      & \ref{fig:w1-loss-new}i
      & --
      \\
      \hline
    \end{tabular}
  \end{minipage}
  \caption{\textbf{Overview of measurements presented in this work.}
    Overview of the nine different waveguide ensembles which were measured. The \emph{Minimum measured propagation loss} has been computed as the lowest point on the graphs referred to in the column \emph{Propagation loss}. The corresponding wavelength is also shown in the adjacent column.
    As an exception to this, the low propagation losses of the W1 waveguides on Sample 2 make the loss at a single point unreliable (see Supplementary Section \ref{seq:w1-losses}), so the table displays a conservative loss (overestimating the loss in comparison to the loss at its single lowest point which is used for the other waveguides).
    }
  \label{table:zigzag-measurements-overview}
\end{table}

\FloatBarrier\section{Imaging the vertically scattered light in straight photonic topological insulator waveguides}
\begin{figure}
  \centering
  \includegraphics{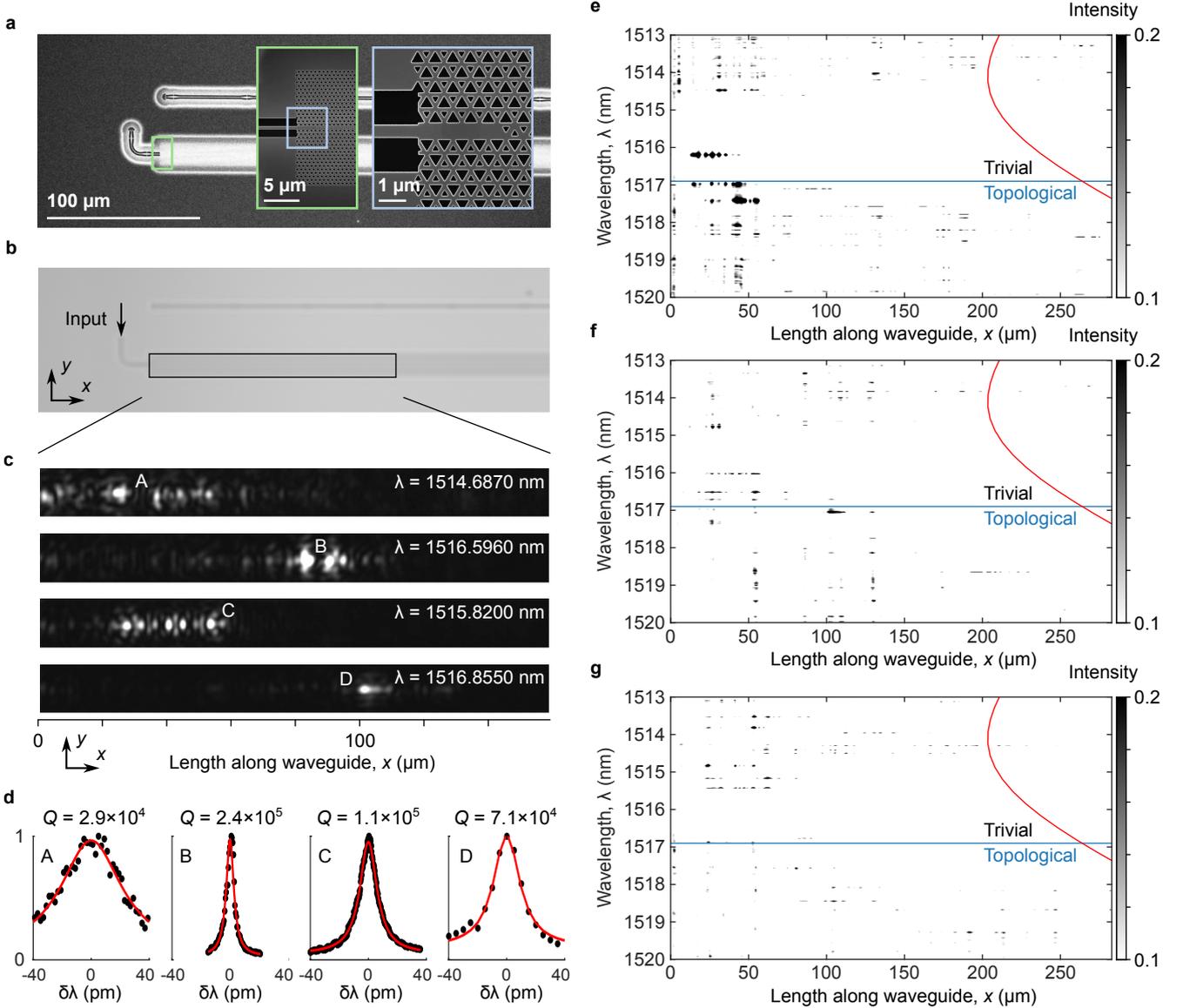}
  \caption{\textbf{Vertically scattered light in straight photonic topological insulator waveguides.}
  \textbf{a}, Scanning electron microscope image of example waveguide (its total length is $1750a_0$). Insets show details of coupling region.
  \textbf{b}, Optical near-infrared image of the example circuit with the input free-space coupler indicated. The black box indicates the rectangle to which the images shown in \textbf{c} are cropped.
  \textbf{c}, Imaged scattered light from the waveguide at different wavelengths showing localized resonances of high $Q$-factor. The modes labeled A and B are from the same circuit, while C and D are from a different circuit.
  \textbf{d}, Frequency dependent intensity of modes imaged in \textbf{c}. The plotted intensities are extracted from a single bright pixel near the labels in \textbf{c}.
  \textbf{e},\textbf{f},\textbf{g}, Wavelength-dependent maps of the imaged intensity on a horizontal line along the waveguide center for the three fabricated devices with total waveguide length $1750a_0$. The red line indicates the loss length, $\ell_L(\lambda)$, plotted as a function of wavelength, found by the analysis described in the main text. The blue line indicates the transition between topological and trivial bands as found by fitting the model given by Eq.~(3) in the main text.
  }
  \label{fig:far-field-straight}
\end{figure}

In addition to the far-field imaging in waveguides including sharp bends, we also perform measurements on the straight waveguides used for the propagation loss extraction (Sample 2). Figure \ref{fig:far-field-straight} shows the emitted light collected along the waveguide axis. Figs. \ref{fig:far-field-straight}c and d evidence the formation of tightly localized and high-$Q$ optical modes at wavelengths around the group index maximum for one particular waveguide of length $L=1750a_0$ (with $a_0=\SI{512}{nm}$). For clarity, the spectro-spatial maps along the waveguide axis for the three nominally identical waveguides are given in Figs. \ref{fig:far-field-straight}e--g. We observe distinct localized optical modes with decreasing extent towards the group index maxima where the propagation losses are shown to peak (see Fig. 3a in the main text and red solid lines included in the maps). The transition from the topological to the trivial branch is also indicated, using the same approach as in the main text. These measurements already evidence strong coherent backscatering leading to complex interference patterns for the topological interface state and even more strongly localized than those shown in Fig. 4 in the main text. However, the exact wavelength for the degeneracy point might vary from waveguide to waveguide (cf. Supplementary Section~\ref{sec:sample-overview}), so we consider that imaging after a sharp bend constitutes a more clear fingerprint of strong backscattering unequivocally occuring for the topological mode, since the bend establishes the topological nature of the transmitted mode (see Section~\ref{sec:bend-simulations}).

\FloatBarrier\section{Modeling light transmittance in sharp-bend geometries\label{sec:bend-simulations}}
\begin{figure}
  \centering
  \includegraphics{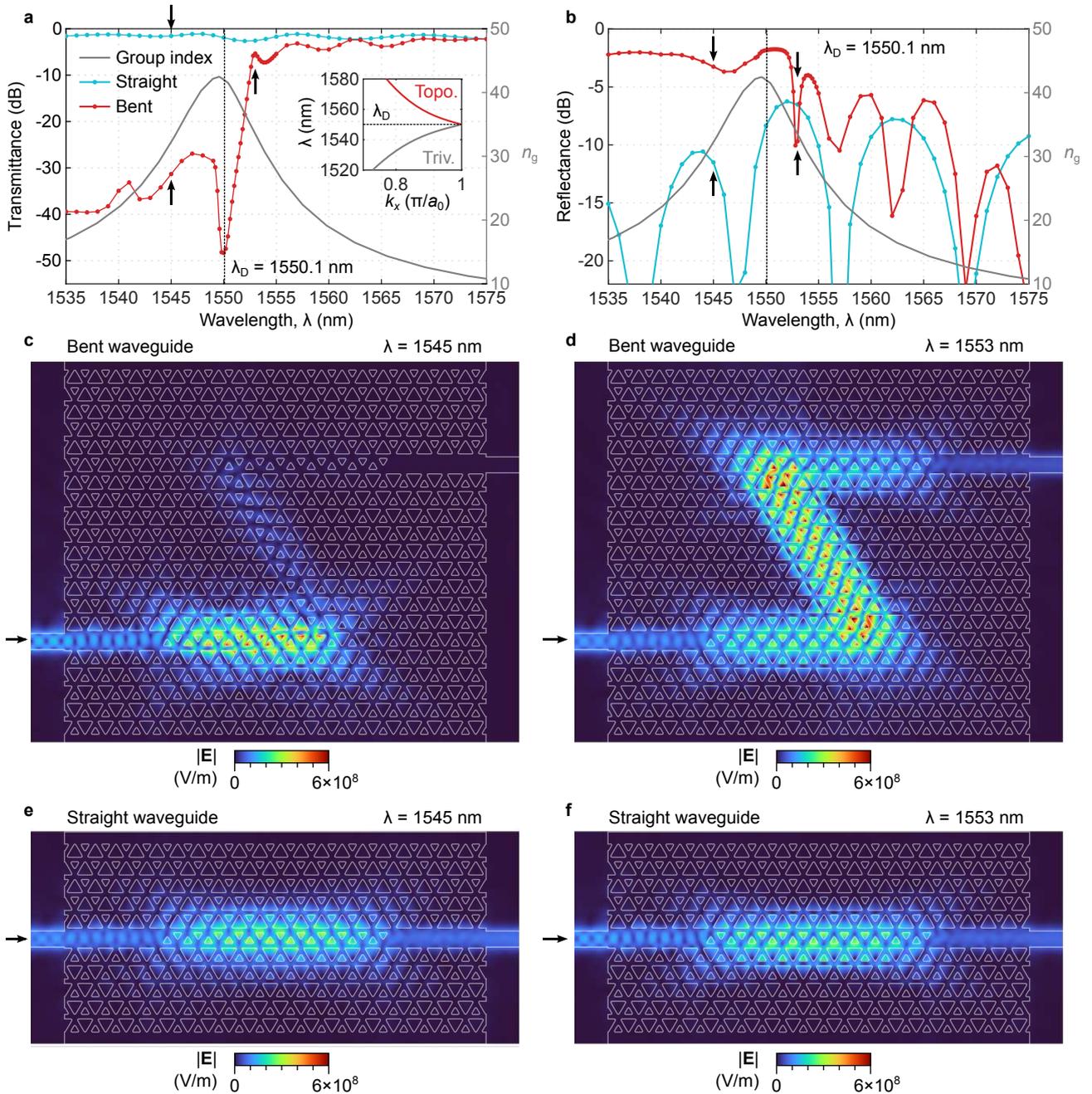}
  \caption{\textbf{Numerical transmittance and reflectance spectra through sharp bends.}
  \textbf{a},\textbf{b}, Transmittance and reflectance spectra of the straight waveguide (cyan) as well as the waveguide with sharp bends (red). The degeneracy point, $\lambda_D$, is marked by a vertical line. The group index, $n_{\mathrm{g}}$, is overlayed in gray on both plots with a scale given by the right vertical axis. The inset of \textbf{a} shows the dispersion of the simulated crystal, where the topological (trivial) band is shown as a red (gray) line.
  \textbf{c}, Simulated geometry and electric field amplitude for the bent waveguide at a wavelength of $\lambda=\SI{1545}{nm}$ corresponding to the trivial band and a group index of $n_{\mathrm{g}} = 31$.
  \textbf{d}, Simulated geometry and electric field amplitude for the bent waveguide at a wavelength of $\lambda=\SI{1553}{nm}$ corresponding to the topological band and a group index of $n_{\mathrm{g}} = 31$.
  \textbf{e}, Simulated geometry and electric field amplitude for the straight waveguide at a wavelength of $\lambda=\SI{1545}{nm}$.
  \textbf{f}, Simulated geometry and electric field amplitude for the straight waveguide at a wavelength of $\lambda=\SI{1553}{nm}$.
  \textbf{c},\textbf{d},\textbf{e},\textbf{f}, The displayed field strength assumes an injected power of \SI{1}{W}.
  \label{fig:bend-simulations}}
\end{figure}

We apply three-dimensional frequency-domain finite-element modelling to explore the transmission properties of PTI waveguides with sharp \SI{120}{\degree}-bends. Similar to the modeling described in Supplementary Section~\ref{sec:insertion-loss}, we simulate the fabricated geometry using a slightly simplified version of the outlines traced for Sample~1 (see Supplementary Section~\ref{sec:sem-trace}; the inset of Fig.~\ref{fig:bend-simulations}a shows dispersion relation) and a short waveguide of length $11.5a_0$. We also simulate a comparable structure where two \SI{120}{\degree}-turns have been introduced inside the PTI crystal to offset the output waveguide interface by $10$ lattice periods compared to its input $y$-location as in the fabricated structure. The transmittance and reflectance of the straight and bent waveguides are shown in Fig.~\ref{fig:bend-simulations}a and b. We observe that the transmittance drops dramatically just before the interface transitions from topological to trivial, with a \SI{25}{dB} extinction occuring within only \SI{3}{nm}. Importantly, accross these \SI{3}{nm}, we observe a considerable drop of the transmittance with the group index value inside the topological mode, as already hinted at in Ref.~\cite{Xie_TopologicalCavityBased_2021}. This clearly confirms that the vertically scattered light seen after the sharp bend in Fig. 5 in the main text results from the topological mode. The simulated reflectance of both the straight and bent waveguide exhibit fringes associated to reflections at the different facets, although these are suppressed within the trivial band for the bent waveguide. This, in addition to the absolute reflectance value (\SI{-2.8}{dB}), indicates that a considerable amount of light is vertically scattered at the first corner. Simulated geometry and electric field amplitude for the bent waveguide at wavelengths in the trivial and topological band exhibiting comparable group indices of $n_g = 31$ are shown in Fig.~\ref{fig:bend-simulations}c and d. Fig.~\ref{fig:bend-simulations}e and f show field amplitude for the straight waveguide at the same wavelengths, where the transmission is large and comparable for both modes.

\FloatBarrier\section{Unraveling the topological and trivial nature of the two modes}
\begin{figure}
  \centering
  \includegraphics{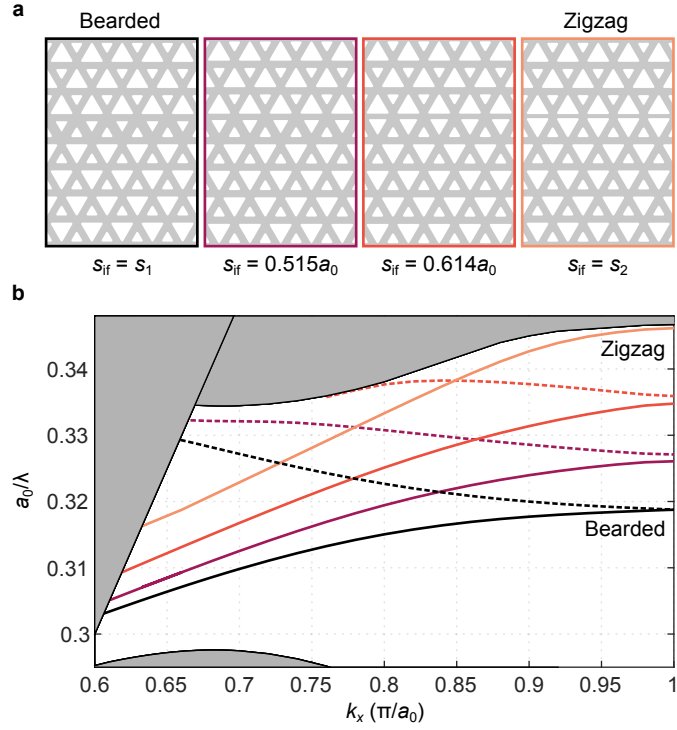}
  \caption{\textbf{Continuous transition between zigzag and bearded interface.} \textbf{a}, Visualization of the geometry of the crystal interface as it is continuously deformed (left to right) from
    $s_\text{if} = s_2 = 1.3a_0/\sqrt3 = 0.750a_0$
    over
    $s_\text{if}=0.614a_0$ and
    $s_\text{if}=0.514a_0$,
    to
    $s_\text{if} = s_1 = 0.7a_0/\sqrt3=0.404a_0$.
    \textbf{b}, Dispersions computed for the zigzag (beige lines) and bearded (black lines), as well as the intermediate interfaces shown in \textbf{a} (purple and orange lines). The additional glide-plane symmetry of the bearded interface enforces a degeneracy at $k_x = \pi/a_0$, which is lifted when $s_\text{if}\neq s_1$. Under the deformation, the trivial band (dashed lines) shifts into the bulk bands whereas the topological band (solid lines) remains outside.
    \label{fig:interface-deformation}
  }
\end{figure}

As described in the main text, the proposed waveguide design supports two guided modes within the band gap of the bounding PTIs. In accordance with the bulk-edge correspondence theorem for systems exhibiting time-reversal symmetry, the difference between the number of forward and backward-propagating topological interface states is equal to the difference between the valley-Chern numbers of the corresponding energy levels for the unit-cell bands of each PTI \cite{Mong_EdgeStatesBulkboundary_2011,Ma_AllSiValleyHallPhotonic_2016}. This indicates that only one of the two guided modes is topological. In Fig.~1 in the main text, we identify the topological mode via transmission across Z-shaped bends (see also Supplementary Sections~\ref{sec:bend-simulations} and \ref{sec:sample1-analysis}). We independently confirm this distinction by relying on the underlying properties of topological modes, which are preserved under continuous deformations of the associated differential operator, as long as the bandgap remains open. In Ref.~\cite{Yoshimi_SlowLightWaveguides_2020}, it was demonstrated that the trivial mode disappears under a continous transformation to a different topological interface \cite{shalaev_robust_2019} while the topological mode is kept intact.
Since our structures differs slightly in refractive index and geometry, we repeat the calculation with a geometry more closely matching that used in our experiments, and reach the same conclusion: The deformation, which changes the size of a single triangle in the unit cell, is displayed in Fig.~\ref{fig:interface-deformation} and causes the trivial mode to merge with the upper bulk while the topological mode remains between the bulk bands. This confirms that only the latter is topological. 

\FloatBarrier\section{Analysis of the effect of fabrication-induced rounding}

\begin{figure}
  \centering
  \includegraphics{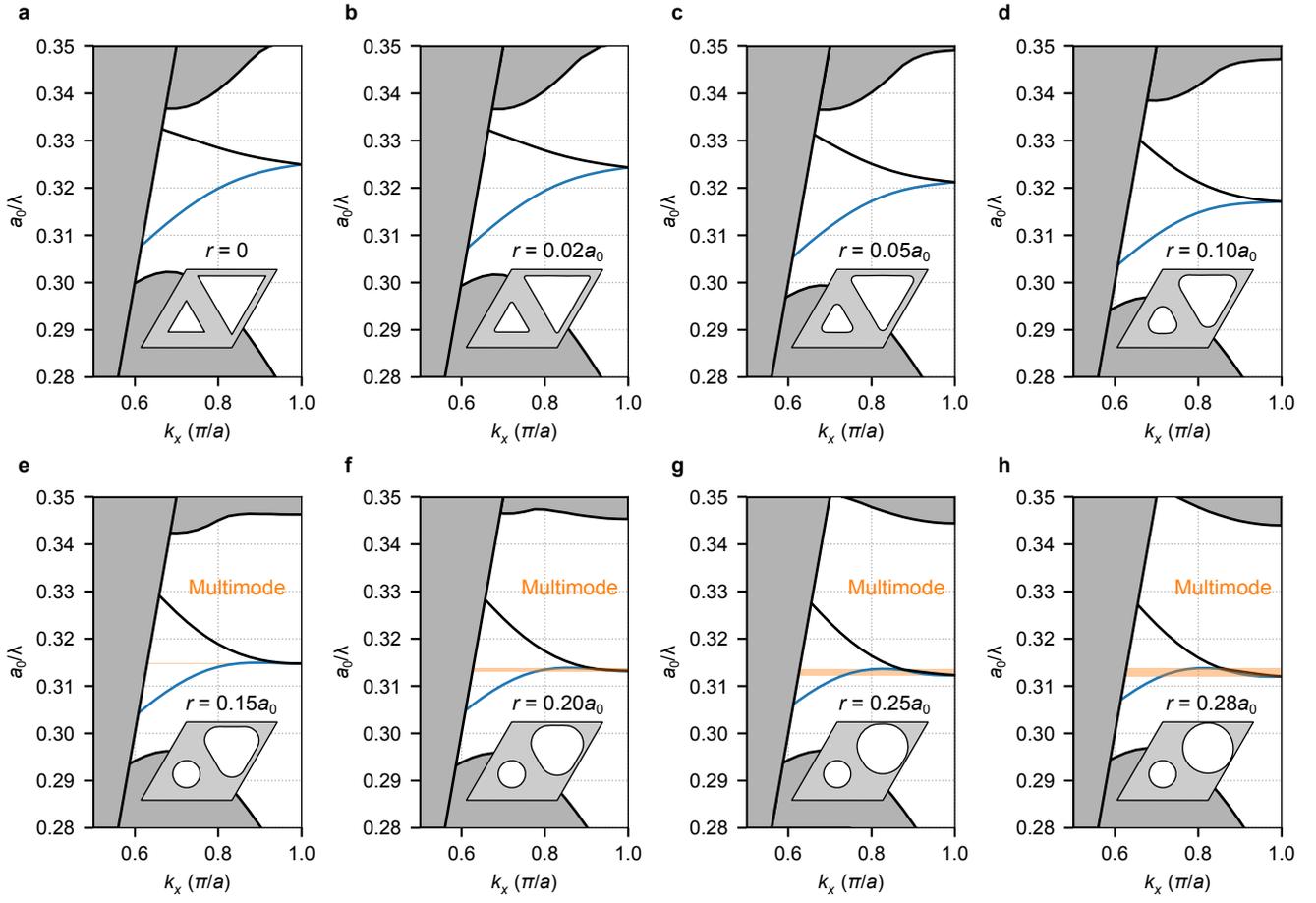}
  \caption{\textbf{Dispersion diagram under transition from triangular to circular holes.}
    \textbf{a}--\textbf{h} Dispersion diagrams for a bearded interface constructed with progressively rounded holes. The unit unit cell is shown in the respective inset and the radius of curvature, $r$, introduced for the triangle corners is indicated above. \textbf{a}, Dispersion diagram and unit cell with air holes that are perfect triangles with sharp corners. \textbf{c}, Dispersion diagram and unit cell with $r=0.05a_0$ which most closely approximates the unit cells of the fabricated structure of this manuscript. \textbf{e}, Dispersion diagram and unit cell with $r=0.15a_0$ which is the point where the smallest hole becomes a perfect circle of the same radius. Hereafter, the parameter $r$ no longer applies to the small hole. This is also approximately the point where the structure becomes multi-moded. \textbf{h}, Dispersion diagram and unit cell with $r=0.28a_0$ which describes the point at which the largest hole has fully transitioned to a circle. Among the structures shown here, this geometry also exhibits the widest spectral band of multi-moded behavior.
  \label{fig:triangle-circle-transition}
  }
\end{figure}

The literature on VH waveguides contains several structures constructed from a unit cell which uses circular holes~\cite{he_silicon--insulator_2019,JalaliMehrabad_ChiralTopologicalPhotonics_2020,hauff_chiral_2022} rather than triangular holes as in our case. Using circular holes is compelling since sharp corners can pose a challenge during fabrication and sometimes require, e.g., bespoke shape correction in the lithography \cite{Barik_TopologicalQuantumOptics_2018}. In any case, sharp corners will be rounded in any fabricated structure. In our case, the rounding is approximately described by a radius of curvature of \SI{23}{nm} \cite{Albrechtsen_NanometerscalePhotonConfinement_2021a}). For this reason, all topological waveguide dispersion diagrams (apart from those in this section) are computed from average outlines extracted from SEM images on a per-sample basis and thus take rounding, as well as other average fabrication-induced deformations, into account.

The main issue with existing VH waveguides based on circular holes~\cite{he_silicon--insulator_2019,JalaliMehrabad_ChiralTopologicalPhotonics_2020,hauff_chiral_2022} is that they exhibit multi-moded behavior adjacent to spectral regions with slow light. This makes them unsuitable for our analysis, since intermodal scattering complicates the analysis. To ensure the single-modedness of our geometry, we consider an interface built from a unit cell that smoothly transitions between exact triangles and circles. The transition is implemented by introducing, at the triangle vertices, fillets that grow until both triangles have fully transitioned to circles. Figure~\ref{fig:triangle-circle-transition} shows the evolution of the interface dispersion during this transition. The filling factor is kept constant during this transition. Moreover, the geometry is chosen to closely resemble the fabricated geometry of Sample 2 at the point where the radius of curvature is about \SI{23}{nm}, corresponding to a radius of curvature of around $r=0.05a_0$ where $a_0=\SI{512}{nm}$. We see that the waveguide is single-moded and remains so until a radius of curvature of about $r=0.15a_0$ (Fig.~\ref{fig:triangle-circle-transition}e), which is about three times greater than our radius of curvature.

\FloatBarrier

\clearpage
\printbibliography
\renewcommand{\thepage}{S\arabic{page}} 

  \renewcommand{\thepage}{S\arabic{page}} 
  \clearpage
\end{refsection}

\end{document}